\documentclass{aa}  

\usepackage{graphicx}
\usepackage[varg]{txfonts}
\usepackage{natbib}
\usepackage{color}
\def\rr{$R^{\star}$}
\def\rtr{$R_{\rm tr}$}

\begin{document}
   \title{What will blue compact dwarf galaxies evolve into?}

   \author{Hagen T.\ Meyer
          \inst{1}
          \and
          Thorsten Lisker\inst{1}
          \and
          Joachim Janz\inst{1,2}
	\and
          Polychronis Papaderos\inst{3}
          }

   \institute{(1) Astronomisches Rechen-Institut, Zentrum f\"ur Astronomie der
  Universit\"at Heidelberg, M\"onchhofstra\ss e 12-14, 69120
  Heidelberg, Germany\\
              \email{HTM@x-astro.net,TL@x-astro.net}\\
              (2) Division of Astronomy, Department of Physics, P.O. Box 3000, FI-90014 University of Oulu, Finland\\
              (3) Centro de Astrof{\'\i}sica and Faculdade de Ci\^encias, Universidade do Porto,
				  Rua das Estrelas, 4150-762 Porto, Portugal
             }

   \date{Received 06 November 2012 / Accepted 30 October 2013}

  \abstract{
   We present and analyse the photometric properties of {a nearly
     complete sample of} blue compact dwarf (BCD) and irregular
   galaxies in the Virgo cluster from multi-band SDSS images. Our study intends to
  shed light on the ongoing debate of whether {a structural} evolution
  from present-day star-forming dwarf galaxies in a cluster
  environment into ordinary early-type dwarf galaxies is possible
  based on the structural properties.\\ 
  For this purpose, we {decompose the surface brightness
    profiles of the BCDs into
    the luminosity} contribution of the starburst {component
    and that of their} underlying low surface brightness (LSB)
  {host}. The latter dominates the stellar mass of the BCD. We find
  that the LSB-components of the Virgo BCDs are structurally
  {compatible with} the more compact half of the Virgo early-type
  dwarfs, except for a few extreme {BCDs}. Thus, {after termination of
    starburst activity, the BCDs will presumably fade into galaxies
    that are structurally similar to} ordinary early-type dwarfs. In
  contrast, the irregulars are more diffuse than the BCDs and are
  structurally similar to the more diffuse half of the Virgo
  early-type dwarfs. Therefore, the present-day Virgo irregulars are
  not simply non-starbursting BCDs.\\
  If {starbursts in cluster BCDs are transient phenomena with a
    duration of $\sim$100 Myr or less, during which the galaxies could
    not travel more than $\sim$100 kpc, then a substantial number of
    non-starbursting counterparts of these systems must populate the
    same spatial volume, namely the Virgo cluster outskirts.} {The
    majority of them would have to be early-type dwarfs, based on the
    abundance of different galaxy types with similar colours and
    structural parameters to the LSB-components of the BCDs. However,
    most Virgo BCDs have redder LSB-host colours and a less prominent
    starburst than typical field BCDs, preventing a robust conclusion
    on possible oscillations between BCDs and early-type dwarfs.}
}

   \maketitle
%

\section{Introduction}
\label{sec:intro}
%
Galaxy clusters like the Virgo cluster are characterised by a wide variety in galaxy morphology. 
This morphological variety depends on the local galaxy density, and therefore shows a clear trend
with the clustercentric {distance}, {as was first revealed by \citet{Dressler80} and confirmed by many subsequent studies \citep[e.g.][]{Binggeli90,Jerjen12}.}
At low {clustercentric distances} the dominant galaxy types are early-type galaxies (ETGs): 
elliptical (E) and lenticular (S0), {dwarf elliptical (dE)},
dwarf lenticular (dS0), and dwarf spheroidal (dSph) galaxies, spanning a range from high to low luminosities.
Traditionally, ETGs are associated with a smooth, regular appearance and no
signs of star formation. However, {the structural characteristics of
  early-type dwarfs have recently been shown to be unexpectedly complex \citep{Janz12,Janz13}. Furthermore, they comprise a particular subclass}
of galaxies with relatively blue cores, which indicate recent or still ongoing {low-level} star formation
at very low star formation rates \citep[e.g.][]{Lisker06b}.

Studies of ETGs have shown that they {have the oldest luminosity-weighted stellar ages among galaxies in clusters, resulting 
in red colours}.
On the other hand, dynamically young galaxy clusters (e.g.\ the Virgo galaxy cluster) also contain late-type galaxies, 
which in contrast to the ETGs, are located at greater clustercentric distances. 
At low luminosities, these {comprise} star-forming dwarf galaxies such
as
blue compact dwarfs (BCDs) and irregulars.
As shown by \cite{Vilchez95}, the H$\beta$ equivalent width (EW) of late-type galaxies in the Virgo cluster increases with 
increasing {distance from the center}, echoeing the strong impact of the cluster environment
on galaxy evolution. For this and other reasons, it has become a
common hypothesis that late-type galaxies may be environmentally transformed to early-type dwarf galaxies \citep[e.g.][]{Kormendy85,Boselli06a}.

Tempting questions in this respect are i) how could {the} morphology-density relation be explained, 
ii) which mechanisms are responsible for the gradual transformation of late-type galaxies into ETGs 
{as one approaches the cluster center}, and iii) are the descendants of today's {low-mass} late-type galaxies {structurally compatible with early-type dwarfs in the Virgo cluster?}
In the literature several mechanisms have been proposed as drivers of galaxy transformations within the dense cluster environment, most notably \emph{ram pressure stripping} \citep{Gunn72}, \emph{tidal stirring} \citep{Mayer01} and \emph{harassment} \citep{Moore96, Mastropietro05}, all of which have the removal of gas in common 
\citep[see e.g.][for a recent review]{Hensler2012-JENAM}.

Dwarf galaxies with a low central stellar density $\rho_{\star}$, such as irregulars, are expected to be particularly prone to 
gas removal when they plunge into the hot intracluster medium (ICM), whereas high-$\rho_{\star}$ systems, such as BCDs \citep[e.g.][]{Papaderos96a}, might be able to retain some fraction of their gaseous reservoir down to a lower {clustercentric distance}.
But what would these dwarfish late-type galaxies in the cluster periphery 
look like after some billion years, once their gas has been removed, in the course of one or several passages
through the dense ICM core and their ensuing long passive photometric evolution?
Addressing this question is fundamental to understanding the morphological diversity of the dwarf galaxy 
population in clusters. Another issue of special interest is what impact the initial contact of late-type dwarfs 
with the cluster periphery has on their star formation activity and whether, for certain conditions, starbursts can be ignited, transforming 
them into BCDs. If so, then how do these cluster-BCDs differ from the main population of field BCDs in their 
recent star formation history and the morphological properties of their star-forming component?

BCDs
are low-luminosity galaxies 
(${M_B > -18 \; {\rm mag}}$) with a compact optical appearance and blue integral colours \citep{ThuanMartin81}. 
Many studies over the past decades have shown that BCDs are
metal deficient with a median oxygen abundance of ${12+\log(O/H) \sim 8.0}$ \citep[see e.g.][for a compilation of literature data]{Kunth00} 
and a low percentage of systems with a gas-phase metallicity as low as ${7.0 \lesssim 12+\log(O/H) \lesssim 7.6}$
\citep{Searle72, Izotov99a, Kunth00, Kniazev04, Papaderos08}. 
{Various lines of evidence consistently suggest that} these systems (XBCDs\footnote{The X indicates the \emph{extreme} properties of these BCDs.}) {have formed most of their stellar mass at late cosmic epochs. They} are therefore the best nearby analogues of young low-mass galaxies in the
early Universe \citep[see e.g.\ the discussion in][]{Papaderos08}. 
Apart from their low metallicity, BCDs also exhibit strong bursts of star formation, which 
are fed by a relatively large amount of gas \citep{ThuanMartin81, Staveley-Smith92, Zee98b}. 
Various arguments suggest, in line with evolutionary synthesis models, that starbursts in {typical field} BCDs do not 
last longer than $\sim 10^7$ yr  \citep{Thuan91, Krueger95, Mas-Hesse99, Thornley00} 
and have to be separated by long ($\sim$1 Gyr) quiescent phases.
Dwarf irregulars, on the other hand, are characterised by prolonged low-level star formation   
{that lasts} 450 Myr up to 1.3 Gyr, as shown by, e.g., \citet{McQuinn10b} through colour-magnitude studies of 20 of these systems.

The seminal study by \citet{LooseThuan86bcdscheme} has shown that BCDs are composed of two main stellar
components. The first one comprises the region of the ongoing star formation where young OB-stars {dominate, and which therefore has} a low {stellar mass-to-light ratio (M/L)}. 
Due to the simultaneous formation of new massive stars in the starburst and their death within {a few} Myr, metal-enriched galactic winds {driven by} multiple supernova explosions are expected to influence the chemical evolution of BCDs. 
However, studies have shown that in most cases starburst-driven feedback is insufficient for the expulsion of 
the entire ISM from a BCD \citep[e.g.][]{Silich98, Ferrara00, Tajiri02, Recchi06}.

The starburst component contributes, on average, $\sim$50\% of the optical emission of a BCD 
and, in some cases, up to $90$\% \citep{Papaderos96a, Noeske99, Cairos01a, Amorin09}. 
Quite importantly, in several XBCDs nebular emission has been determined to be extraordinarily intense
(${EW \sim 1600-2000}$\,\AA) and to contribute 30--50\% of their total optical luminosity \citep{Papaderos98a, Papaderos02},
in line with theoretical predictions for young starbursts \citep[e.g.][]{Krueger95}.

The second component of BCDs is dominated by an old population of {low-mass} stars. 
This component is referred to in the following as the \emph{host galaxy} or \emph{low-surface brightness (LSB)}\footnote{Note: The LSB-component of BCDs should not be mixed up with the galaxy type commonly called ``Low Surface Brightness galaxies''.
To the contrary, a high central surface brightness ($\mu_{0}(B)<$22 mag/arcsec$^{2}$) has been found to be a characteristic 
property of the LSB-host of BCDs \citep{Papaderos96b,GildePaz05} implying a higher central mass density than in 
{typical} irregulars, and the more so, in dSphs and genuine ($\mu_0 \geq 23.5$ mag/arcsec$^{2}$) LSB galaxies.} 
component and is characterised by a high stellar M/L. 
Various studies indicate that the LSB-component dominates the baryonic-, and in some cases even the virial mass of BCDs within their Holmberg radius.\footnote{The Holmberg radius defines the radius where $\mu(B) = 26.5$ mag/arcsec$^2$.}
 Therefore, it has to have a significant influence on the gas collapse characteristics and the starburst activity in these systems \citep[e.g.][]{Papaderos96b, Oestlin99, Elmegreen12, Lelli12, Micheva13}.

\begin{table}
\caption{Different subtypes of BCDs according to the morphology of the starburst- and LSB-component based on \citet{LooseThuan86bcdscheme}.
}\tabularnewline         
\label{BCDsubtypes}      
\centering          
\begin{tabular}{cl}
\hline\hline       
Subtype	&	Description	\\\hline
nE		&	nuclear star-forming region and\\ 
		&	elliptical LSB-component\\
iE		&	irregular star-forming region(s)\\
		&	and elliptical LSB-component\\
iI		&	irregular star-forming region(s)\\ 
		&	and LSB-component\\
iI,C		&	iI with cometary shape\\ 
i0		&	no detected LSB-component\\
\hline                  
\end{tabular}
\end{table}

According to \citet{LooseThuan86bcdscheme,Loose86}, BCDs can be classified in four main {subtypes}, based on the morphology of their 
star-forming and  LSB-component {(Table~\ref{BCDsubtypes}; see \citealt{Salzer89b}, \citealt{Telles97}, \citealt{Cairos01b} and \citealt{Sung02} for alternative classification schemes.)}.
Nuclear-elliptical (nE) and irregular-elliptical (iE) BCDs {are} both characterised by an extended circular or 
elliptical LSB-component, {while BCDs with an irregular LSB-component are classified ``iI''. A recent study by \citet{Zhao13} found that iI-type BCDs are on average more gas-rich and metal-poor than nE/iE-type BCDs.}
Interestingly, the {\sl cometary} (iI,C) {subtype} \citep[see e.g.][for two nearby examples]{Noeske00} is  
remarkably common among XBCDs \citep{Papaderos08}, a fact pointing to a connection between 
gas-phase metallicity, morphology and evolutionary status.
This is also indirectly suggested by the high frequency of cometary galaxies (also referred to as tadpoles) 
among comparatively unevolved high-z galaxies \citep[e.g.][]{Elmegreen10}.

Cometary morphology in field XBCDs has been proposed to result from 
unidirectional sequential star formation activity with a typical velocity of the sound speed in the warm ISM \citep[$\sim$20 km/sec, ][]{Papaderos98a, Papaderos08}. 
This scenario is supported by stellar age gradients along the ``comet's tail'' \citep[e.g.][]{Guseva03}. 
In the cluster periphery, however, cometary morphology may arise from extranuclear star formation that 
is triggered through the interaction with the ICM 
\citep[see][for a recent review]{Hensler2012-JENAM}, as, e.g.\ the impressive cases of two star-forming dwarf galaxies 
in Abell~1367 \citep{Gavazzi01} demonstrate.

The evolution of star-forming dwarf galaxies like BCDs to ``red and dead'' early-type galaxies is still under debate and there is 
no satisfying answer yet. 
Early studies by \citet{Thuan85} and \citet{Davies88} introduced the {idea of a} possible evolution of irregulars to BCDs in several bursts and finally, after reaching a 
higher metallicity and the depletion of gas, the fading to early-type dwarfs. \citet{Thuan85} concluded that the metallicities of BCDs and early-type dwarfs are very different 
and that only periods of three to ten bursts over the Hubble time could be able to produce the metallicity range of early-type dwarfs.
The study of \citet{Marquart07} on the BCD He 2-10 showed that the stars in this galaxy have random motions and show signatures of a merger.
They concluded that, due to the velocity dispersion of the stars, in the future this BCD may evolve into a nucleated early-type dwarf. In comparison to this, the simulations of \citet{Bekki08} showed that dwarf-dwarf mergers are able to produce BCDs, but the further evolution of these BCDs into gas-free dwarfs would require additional influence to remove the extended gas discs of the simulated BCDs.

In contrast to BCDs, early-type dwarf galaxies {typically} show almost no
evidence for ongoing star formation. However, several studies (e.g.\ \citealt{Vigroux84}, \citealt{Gu06}, \citealt{Lisker06b}) found early-type dwarfs with blue colours in their central region,
indicating recent or ongoing star formation, partly with post-starburst spectra that may hint at a population of recent  arrivals to the cluster environment \citep{Gavazzi10}.
Among 476 early-type dwarfs in the Virgo cluster, \citet{Lisker06b} identified 23 galaxies with blue centers and classified them as ``dE(bc)''.
Since the debate of the evolutionary connection of early-type dwarfs and BCDs is
still ongoing, it is
worth comparing the structural properties of
  BCDs to those of early-type dwarfs and in particular to those of dE(bc)s.
Can BCDs in the periphery of the Virgo cluster evolve into objects similar to the present-day early-type dwarfs? Would this be different for field BCDs?

This paper is structured as follows: Section 2 defines our sample of cluster BCDs and the 
classification criteria used.
In Section 3 we outline our photometric measurements, with particular emphasis on the decomposition methodology for BCDs. Our results on the structural properties of BCDs and the comparison to early-type dwarfs and irregulars are presented in Section 4. In Section 5, we discuss the results and conclude with an assessment of possible evolutionary connections between the different dwarf types.


\section{Sample selection and SDSS data}
\label{sec:sample}
Our sample is based on the Virgo Cluster Catalog (VCC) by
\citet{Binggeli85}, which includes galaxies of all types within the
Virgo cluster area.
Due to incomplete velocity information for the VCC galaxies, the VCC includes certain and possible cluster members \citep[updated by][]{Binggeli93}, apart from background galaxies. Since new velocities have become available in the meantime, largely due to the Sloan Digital Sky Survey \citep[SDSS,][]{Adelman07}, the membership was revised by one of us (T.\ Lisker, see appendix of \citealt{Weinmann11}) using the NASA/IPAC Extragalactic Database\footnote{The NASA/IPAC Extragalactic Database (NED) is operated by the Jet Propulsion Laboratory, California Institute of Technology, under contract with the National Aeronautics and Space Administration.}. If a galaxy is listed as certain or possible member in the VCC, but has a velocity above 3500\,km/s, it is considered as a background galaxy.
We do not change possible to certain members or vice versa.

{Here we consider both }certain and possible members to a magnitude limit 
of ${m_B \le 18.0}$ mag, to which the VCC was found to be complete \citep{Binggeli85}.
When applying a constant distance modulus of ${m - M = 31.09}$ mag \citep[$d = 16.5$ Mpc;][]{Mei07} to all galaxies, this corresponds to an absolute magnitude limit of ${M_B < -13.09}$ mag.

\begin{table}
\caption{{VCC galaxies classified as BCD or candidate BCD.}
\label{all-BCDs}} \tabularnewline
\centering
\begin{tabular}{c l} 
\hline\hline 
Amount & Type\\ \hline
38	&	``BCD'', ``BCD?'', ``BCD:'',``BCD or merger''\\
1	&	``pec'' $\rightarrow$ ``BCD?''\\
10	&	``Im / BCD'', ``Im III / BCD'', ``Im III / BCD:''\\
	&	 ``Im III / BCD?'', ``Im III,pec / BCD''\\
2	&	``?'' $\rightarrow$ ``Im / BCD''\\
3	&	``Spec / BCD'', ``Spec,N / BCD''\\
3	&	``Sm III /  BCD'', ``SBm III / BCD''\\
1	&	``dS? / BCD?''\\
1	&	``Sd / BCD?''\\
1	&	``dS0 or BCD''
\\\hline
\end{tabular}
\tablefoot{
{
The galaxy classification was adopted from the VCC, except where we indicate our new classification with an arrow. In the VCC, an uncertainty in the morphological classification of the galaxy is indicated by ``:'', a {more severe} uncertainty by ``?''. Roman numerals are the luminosity class as given in the VCC.
}
}
\end{table}

\begin{figure}
\centering
\includegraphics[width=\hsize]{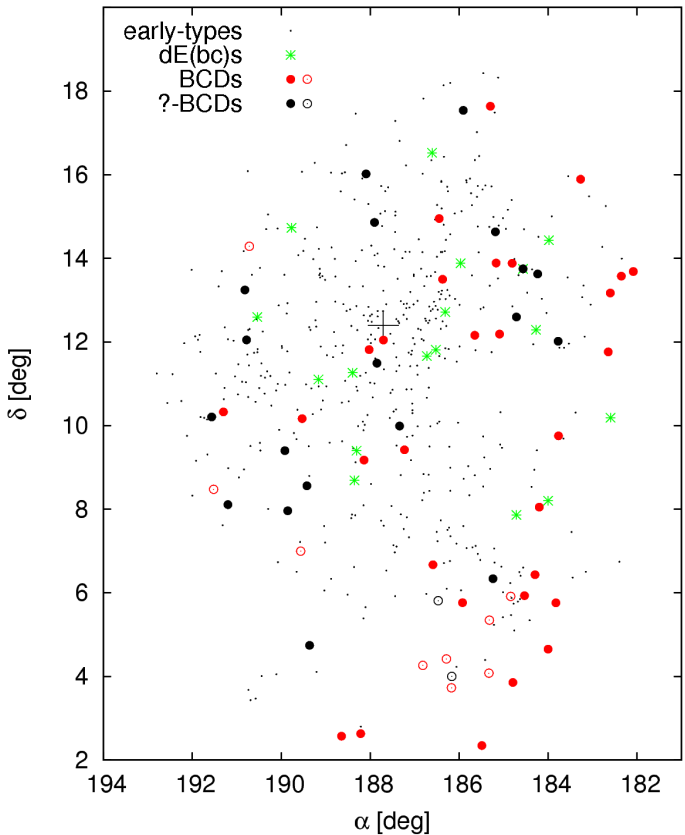}
\caption{{Distribution of Virgo cluster galaxies. Red symbols refer to objects with BCD as primary class, black symbols to uncertain BCD candidates, small black dots to early-type galaxies, green asterisks to dE(bc)s, and the black cross to M87. Open symbols denote galaxies that could not be analysed (see text for details).}
               }
\label{map}
\end{figure}

Within these limits, the VCC contains 57 galaxies which have the term ``BCD'' included in their morphological type, 38 of them with BCD as the only or primary class (Tab.~\ref{all-BCDs}).
Among the galaxies 
whose class was ``?'' and ``pec'' in the VCC, a visual inspection of the SDSS images\footnote{Using the online Image List Tool of the SDSS, http://skyserver.sdss3.org/dr8/en/tools/chart/list.asp\,.}
suggests a reclassification of VCC~0429 and VCC~1713 to the mixed type ``Im / BCD'', and of VCC~1411 to a possible ``BCD?'', thus adding up to 60 galaxies, and 39 with BCD as primary class {(Tab.~\ref{all-BCDs})}.
Since the VCC photographic plates were more sensitive in the blue, it needs to be kept in mind that this could have partly influenced the classification, as any underlying red stellar population would appear less prominent in the blue. Therefore,
the galaxies with primary class BCD are chosen as the working sample for our analysis, whereas the uncertain BCD candidates are treated as a separate sample that is compared to the BCDs in Section~\ref{mixedtypes}.  Figure~\ref{map} shows the distribution of both samples within the Virgo cluster.

Five of the 39 BCDs (VCC~0464, 0468, 0741, 0772, and 0985) were not covered 
by the SDSS Data Release Five (DR5, \citealt{Adelman07}), two were
excluded because of a nearby other galaxy (VCC~0340 and 1944), and two
were excluded because of a nearby bright star (VCC~1750 and 2033). Our
primary working sample thus consists of 30 galaxies.  Two of the 21 uncertain BCD candidates (VCC~0737 and 0848) were not covered by the SDSS DR5, leaving us with 19 such objects.

Our photometric study relies on the SDSS DR5 images in u, g, r, i and z, {each} with 
an exposure time of 54 s. 
Due to insufficient sky subtraction of the SDSS pipeline for nearby galaxies of large apparent size \citep{Lisker07}, we used the sky-subtracted images provided by \citet{Lisker07}. All images were flux calibrated 
and corrected for Galactic extinction following \citet{Schlegel98}.

\section{Surface photometry and BCD decomposition}
\label{photometry}


\subsection{Global photometry of late-type and early-type galaxies}
\label{sec:globalphoto}

The analysis of the sample of ETGs (classes E, S0, dS0, and dE, including dE(bc)\,) was performed by \citet{Lisker07} and \citet{Janz08}. In order to obtain global photometric charateristics for the irregulars and for the uncertain BCD candidates, we closely followed the procedure of  \citet{Lisker07}, as outlined below, but using the co-added $g$, $r$, and $i$ images \citep{Lisker06a} to benefit from a higher signal-to-noise ratio. While we also applied this procedure to our primary BCD working sample -- yielding global characteristics that include the light from the starburst region -- we then proceeded to perform a thorough decomposition of their light profile, as outlined in the following sections. We point out that we did not consider it appropriate to apply a similar decomposition to the irregulars, since these are not clear multi-component objects like the BCDs.

We first determined the circular Petrosian radius \citep{Petrosian76}
on the co-added image, using a visual estimate of the galaxy center\footnote{\citet{Lisker07}
  used so-called asymmetry centering for the early-type dwarfs instead.}
that aims to reflect the overall light distribution, not only the
inner light peak(s) that an irregular may exhibit. We then
computed an elliptical shape, i.e.\ position angle and axis ratio, from the image moments of the enclosed light. This is used to determine a ``Petrosian semi-major axis'' $a_{\rm P}$ on elliptical apertures. At $a_{\rm P}$, we let the IRAF task \textit{ellipse} \citep{Jedrzejewski87} determine the isophotal position angle, axis ratio, and centre. With these values, we performed another iteration on deriving $a_{\rm P}$ as well as position angle and axis ratio $b/a$. In this process, contaminating objects that do not belong to the galaxy were masked.

We obtained total $r$ magnitudes from the flux within $2 a_{\rm P}$ on the $r$-band image, which also yielded the values of  the half-light semi-major axis $a_{\rm hl}$ and the average surface brightness $\left< \mu \right>_{\rm{eff}}$ within $a_{\rm hl}$. In our figures we  show the effective radius,
\begin{equation}
 	{R_{\rm eff} = a_{\rm hl} \cdot \sqrt{(b/a)}}.
\end{equation}
Colours were obtained from the flux within $2 a_{\rm P}$ on the images
of the different bands. Photometric errors were derived using
equations (3) and (4) in \citet{Lisker08a}, based on the
signal-to-noise ratio and on estimations of the uncertainties in the sky subtraction and determination of Petrosian radii.


\subsection{BCD surface brightness profiles}

The luminosity contribution of the starburst {in field BCDs} amounts, on average, to
$\sim50 \%$ of the $B$ band emission, and in some extreme cases reaches up to 
$\sim90 \%$ \citep{Papaderos96a, Noeske99, Salzer99, Amorin09}.
On the other hand, the starburst component is almost negligible, in terms of its 
fraction by \emph{mass}, as its {stellar M/L} is several times lower than that of the LSB-component.
An adequate separation of its emission via 1D or 2D decomposition is clearly necessary for 
isolating the emission and studying the structural properties of the host galaxy.
Indeed, the composite surface brightness profile of a BCD holds little insight into the 
photometric structure prior to decomposition. For example, \citet{Papaderos96a} have pointed out that the profile of a nE-type BCD 
can closely resemble the steep inner profile of a {giant} elliptical, whereas a typical feature of iE-type BCDs is an extended \emph{plateau}.
Similarly, the effective radius ${R_{\rm eff}}$ can vary
by up to a factor $\sim$3 depending on the starburst luminosity fraction \citep[cf.][]{Papaderos06a}.

For galaxies with a smooth appearance like elliptical galaxies the easiest way to obtain surface brightness profiles is 
to use elliptical apertures and sum up the enclosed flux. Due to the irregular morphology of BCDs, however, 
such an approach is impractical, on the one hand because the choice of the 'galaxy centre' is subjective and 
wavelength-dependent, and, on the other hand,
because profiles derived in this way show, in the case of iE- and iI-BCDs, discontinuous jumps. In some cases, the 
latter can significantly affect the profile of the LSB-component, thereby biasing studies of the photometric structure of BCDs. 2D axisymmetric models to BCDs other than those falling into the nE class also yield systematic residuals, unless
one carefully masks out and avoids fitting of the star-forming component, as was done in e.g.\ \cite{Amorin09}.

{With these considerations in mind, we therefore} applied method iv of \citet{Papaderos02}, which was also used by \citet{Noeske06} under the name LAZY.
This method, which has as input a set of co-aligned multi-band images of the same pixel scale (SDSS: 0.396 arcsec/pix) 
and point spread function (PSF), does not require a choice for the 'centre' of a galaxy. 
In our study, an average FWHM of the PSF of 4 pix (1.584 arcsec) was applied to smooth the input images 
with a Gaussian convolution kernel.
From these co-aligned images a S/N weighted average image (called
reference frame) is created, which is used to calculate a mask for
each intensity interval ${\Delta I}$ within the range ${I_{\rm min}}$
to ${I_{\rm max}}$. Pixels outside the intensity interval of ${I- \Delta I \le F \le I}$ are set to zero, while pixels within the interval are 
given full consideration.
To apply LAZY, it was necessary to remove contaminating sources by replacing their flux with the median flux of the immediate environment.

In the next step,
the mask of a given object and intensity interval is multiplied by the image in a given band, and the product's total flux is computed.
The corresponding photometric radius \rr\ is calculated as
\begin{equation}
 	{R^{\star} =  \left( \frac{A_I + A_{I-\Delta I}}{2 \pi} \right)^{1/2}\quad ,	
	}
\label{rphoto}
\end{equation}
where ${A_I}$ and ${A_{I-\Delta I}}$ are the areas with intensities above 
${I}$ and ${I-\Delta I}$. 
In the case of multiple star-forming regions the \rr\ derived in this way corresponds to the sum of their area, a concept
which translates into a monotonous increase in radius with decreasing intensity threshold.

   \begin{figure*}
   \centering
   ~\hspace{0.5cm}\includegraphics[width=0.245\hsize]{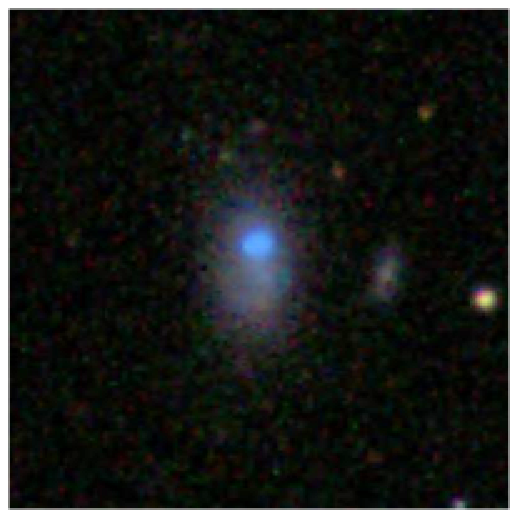}\hspace{3.8cm}
   \includegraphics[width=0.245\hsize]{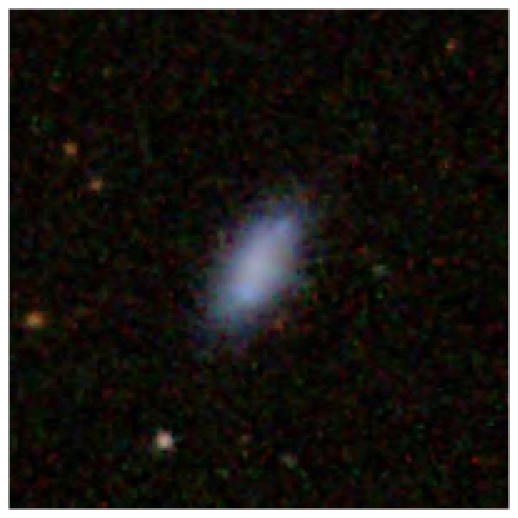}\\
   \includegraphics[width=0.42\hsize]{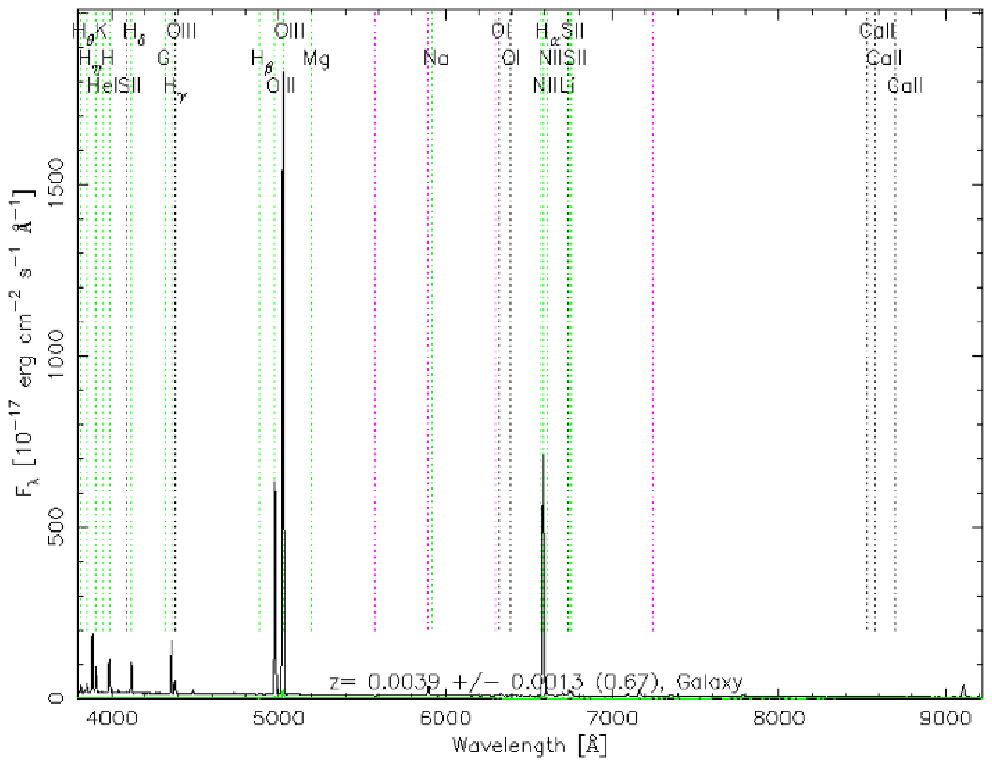}\hspace{5mm}
   \includegraphics[width=0.42\hsize]{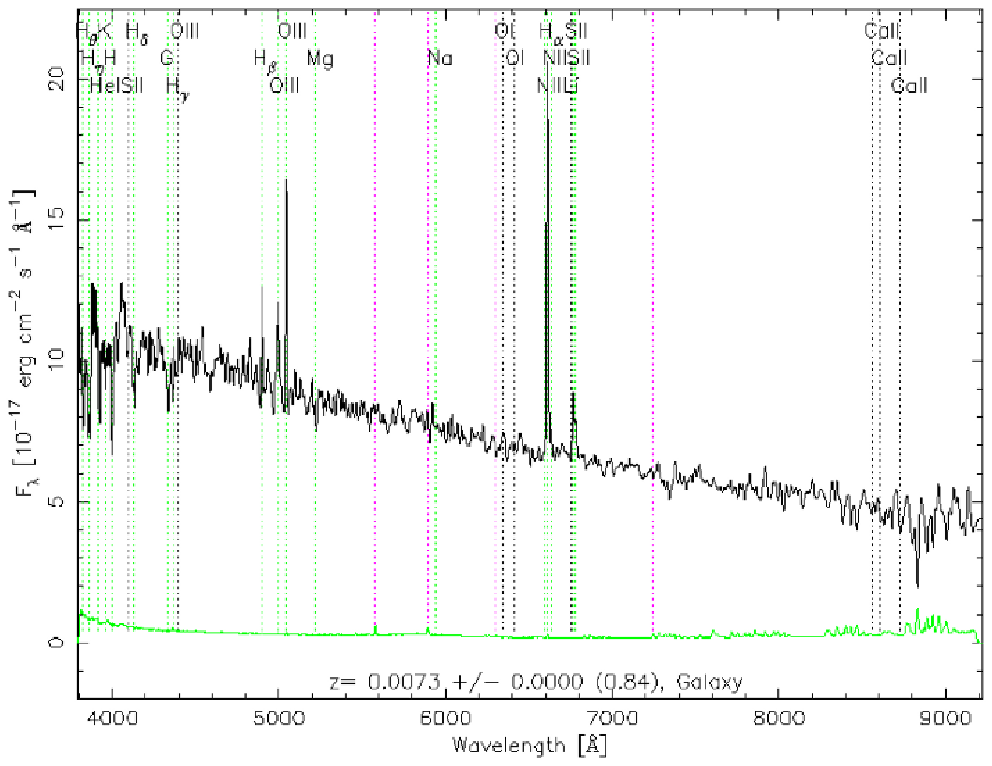}\\
   \includegraphics[width=0.45\hsize]{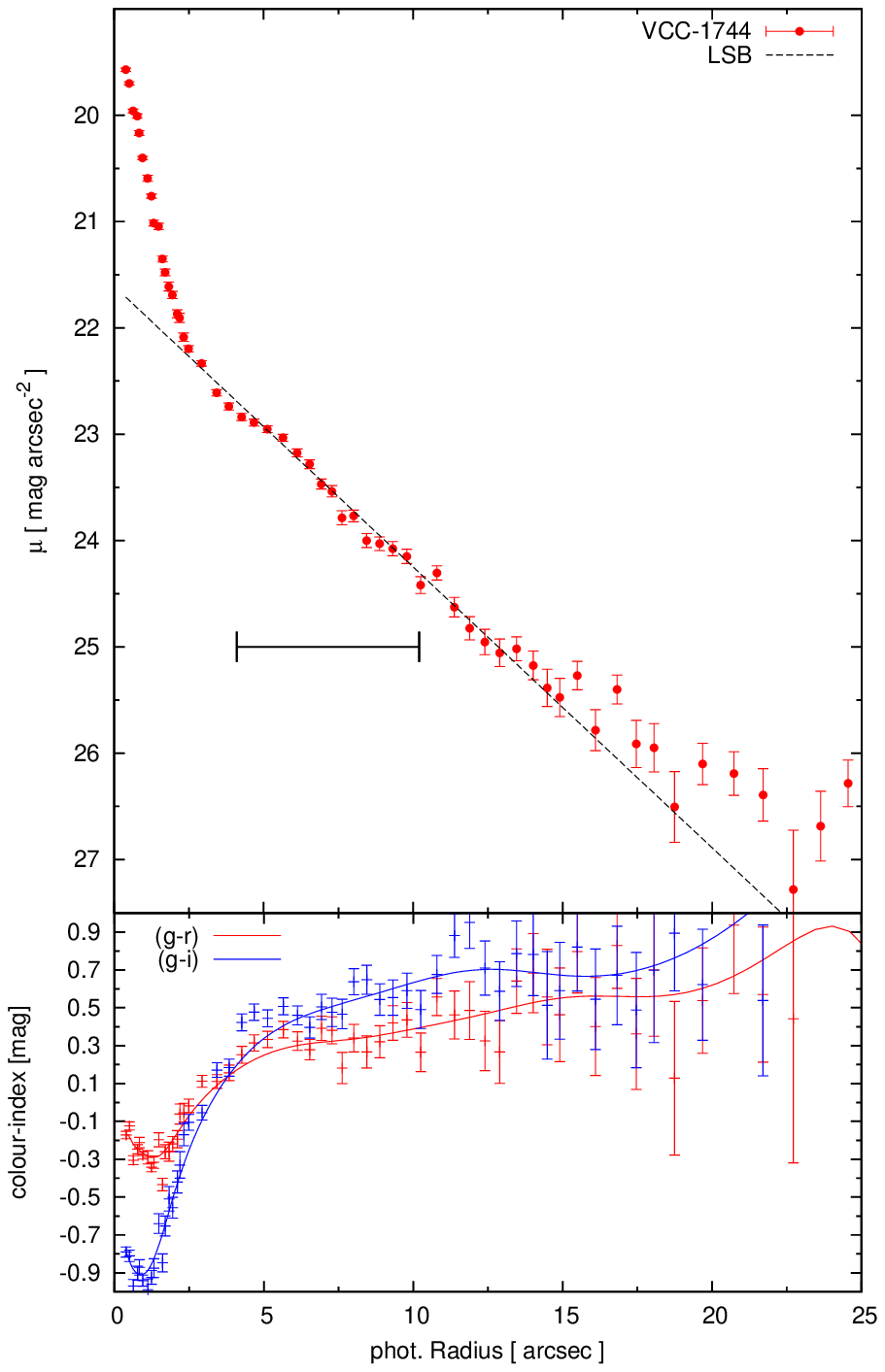}
   \includegraphics[width=0.45\hsize]{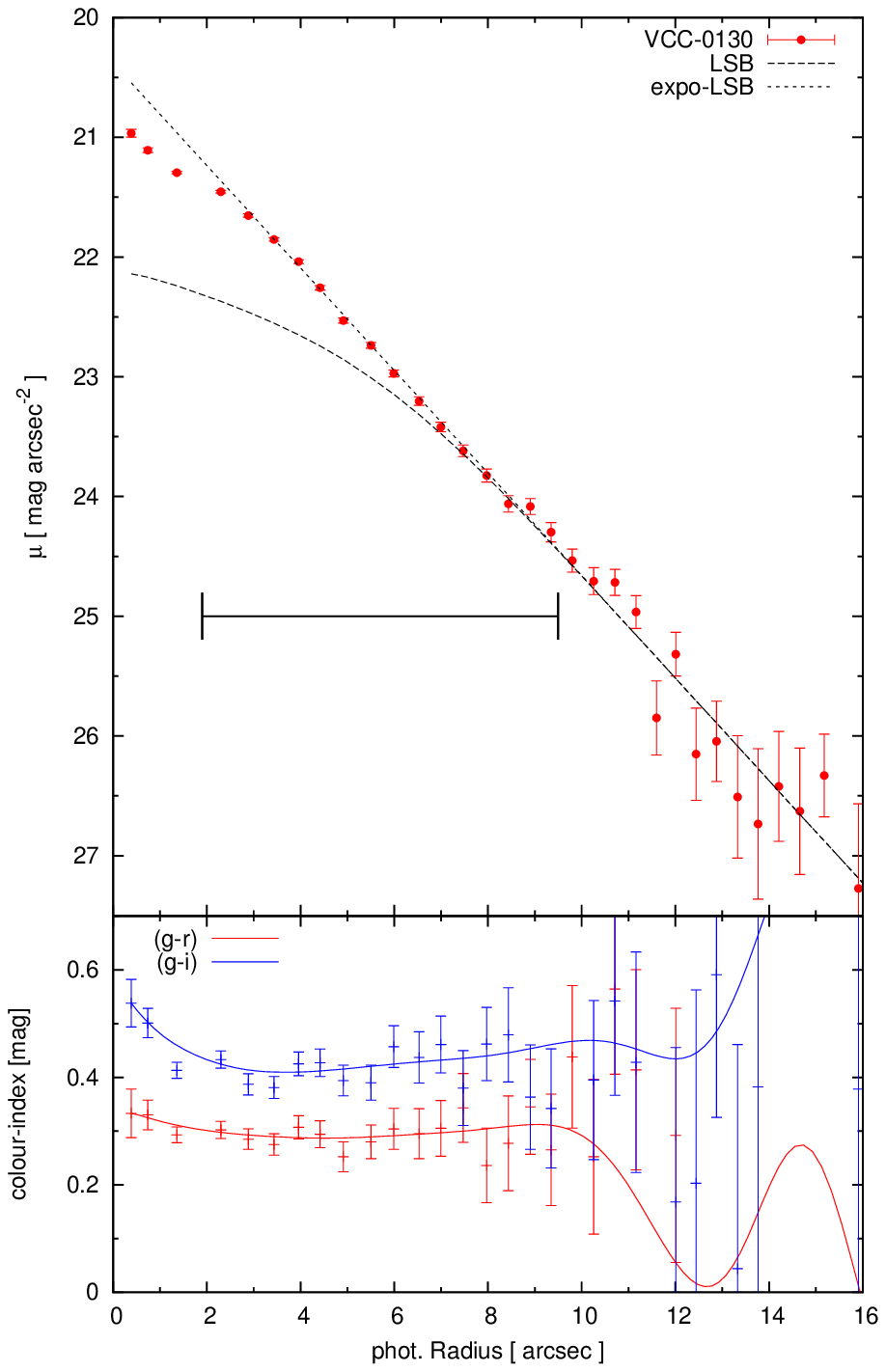}\\
   \caption{{Colour images from SDSS (upper panels), spectra
       (middle panels), as well as surface brightness and colour
       profiles (lower panels) of VCC~1744 (left) and VCC~0130
       (right). In the case of VCC~1744 a pure exponential law was
       assumed to fit the LSB-component, while for VCC~0130 an inner
       profile flattening was assumed  (black dashed lines). The
       horizontal bar denotes the interval in \rr over which the
       LSB-component was fitted.
The images in this figure and throughout the paper have a width of $80"$, corresponding to $6.3$\,kpc at a
distance of $16.5$\,Mpc.
They were obtained from the online Image List Tool of the SDSS (http://skyserver.sdss3.org/dr8/en/tools/chart/list.asp) and were slightly enhanced at low intensity, using the same correction curve for all.
}
}
   \label{fig:examples}
   \end{figure*}

The errors of the data points in the surface brightness profiles (see
e.g.\ Fig.~\ref{fig:examples})
were calculated by {estimating} the Poisson photon noise and
the variation of the background of each SDSS image, {as
  detailed in} \citet{Noeske03a}, as well as in \citet{Papaderos96a} and \citet{Cairos01a}.

\subsection{LSB-component: exponential model}

To distinguish the starburst from  the LSB-component {in the
  galaxies of our primary BCD sample,}  we use (g-i)-colour profiles. 
At smaller \rr\ the (g-i)-colour profile is due to the superposition of the 
contribution of the young population and the old population of the LSB-component, 
which results in relatively blue values.
At larger \rr\ the contribution of the starburst component vanishes and the old LSB-component dominates, which results
in relatively constant red colours.
The radius beyond which the colour index levels off to a red, nearly constant value is referred to as
\emph{transition radius} \rtr\ \citep{Papaderos96a, Cairos03, Noeske03a}. 
We used this characteristic radius to define the minimum \rr\ for fitting an exponential law of the form
 \begin{equation}
 	{I_{\rm LSB}(R^{\star}) = I_{\rm LSB,0} \; exp\left[ -(R^{\star}/\alpha) \right]},
\label{expo}
 \end{equation}
to the LSB-component. In units of mag/arcsec$^{2}$ Equation~\ref{expo} reads as
 \begin{equation}
 	{\mu_{\rm LSB}(R^{\star}) = \mu_{\rm LSB,0} + 1.086 \left( R^{\star}/\alpha \right)},
 \end{equation}
where ${I}_{\rm{LSB,0}}$ and $\mu_{\rm{LSB,0}}$ are the central
intensity and central surface brightness of the LSB-component, and $\alpha$ is its 
exponential scale length.

\subsection{LSB-component: inner flattening}
\label{innerflattening}
In some cases the extrapolation of the exponential fit to \rr=0 arcsec exceeds the intensity observed at intermediate
and small radii. This implies that the exponential law is not applicable in the central part of the LSB-component and must 
flatten to a central surface brightness that is {fainter} than the extrapolated value $\mu_{\rm{LSB,0}}$ of the fit for 
\rr$\geq$\rtr. {This type of outer exponential profile} with a flat core {was} noticed by \citet{Binggeli91} for early-type dwarfs, who have 
called them type~V. \citet{Papaderos96a} introduced a modified exponential law to approximate such profiles, motivated by two considerations: Firstly, a S\'ersic law with $n \lesssim 0.5$ cannot fit type~V profiles without producing systematic residuals (see \citealt{Noeske03a} for a detailed discussion of this subject).
Secondly, as shown by radiation transfer models by \citet{Papaderos96a}, a S\'ersic law with $n \lesssim 0.5$ implies an 
extended 'hole' in the intrinsic luminosity density of a {spherically symmetric} emitter, if radiation isotropy, and a 
uniform M/L ratio and intrinsic extinction are assumed. As these authors considered the evacuation of the dwarf galaxy centres to be improbable, they introduced an intensity profile for which the intrinsic luminosity density
increases monotonously with decreasing radius (for small flattening parameters; see below) and has a finite central value. This modified exponential fitting law (\textit{modexp}) has the form

  \begin{equation}
 	\label{flattening}
 	{I_{\rm{LSB}}(R^{\star}) = I_{\rm{LSB,0}} \; exp \left( -\frac{R^{\star}}{\alpha} \right) \left[ 1 - q \cdot \; exp \left( -P_3(R^{\star})\right) \right] },
 \end{equation}

where
\begin{equation}
 	{P_3(R^{\star}) = \left( \frac{R^{\star}}{b\alpha} \right)^3 + \left( \frac{R^{\star}}{\alpha}  \cdot \frac{1-q}{q} \right)} ,
\end{equation}
with the typical ratio b/q being of the order of three (\citealp{Papaderos99} and \citealp{Papaderos12}). 
The parameter b is a measure of the radial extent of the central core in units of the exponential scale length $\alpha$,
and ${q= \frac{\Delta I}{I_{\rm{LSB,0}}}}$ describes the attenuation of the \textit{modexp} fit, as compared to the 
central intensity ${I}_{\rm{LSB,0}}$ predicted by the pure exponential fitting law.
Due to the poor knowledge of the structure of the LSB-component in its central part 
it is not clear at the moment which parameter combination of (b,q) satisfactorily 
describes the original shape. Therefore, the choice of (b,q) is not a straightforward task and 
has to rely on plausibility considerations \citep[cf.][]{Noeske03a}.
In this study {we chose ${(b,q)} = (2.40\:,0.80)$, by which the total magnitude of the LSB-component becomes $0.48$ mag fainter and $\mu_{\rm{0}}$ is reduced by $1.747 \; \rm{mag/arcsec^2}$ as compared to the purely exponential LSB-component. Since {this magnitude offset} is applied to all filter bands, the {integrated} colours of the LSB-components are not affected. We apply such a flattening only} when $\mu_{\rm{LSB,0}}$ 
exceeds the {total} central surface brightness $\mu_{\rm{tot,0}}$ and the spectrum shows 
clear signs of star formation. This is {only the case for three BCDs (VCC~0074, 0130, and 0641), and is
 exemplified in Fig.~\ref{fig:examples} for VCC~0130.}

\subsection{LSB-component: outer tail}
\label{tail-section}

   \begin{figure*}
   \centering
   \includegraphics[width=0.4\hsize]{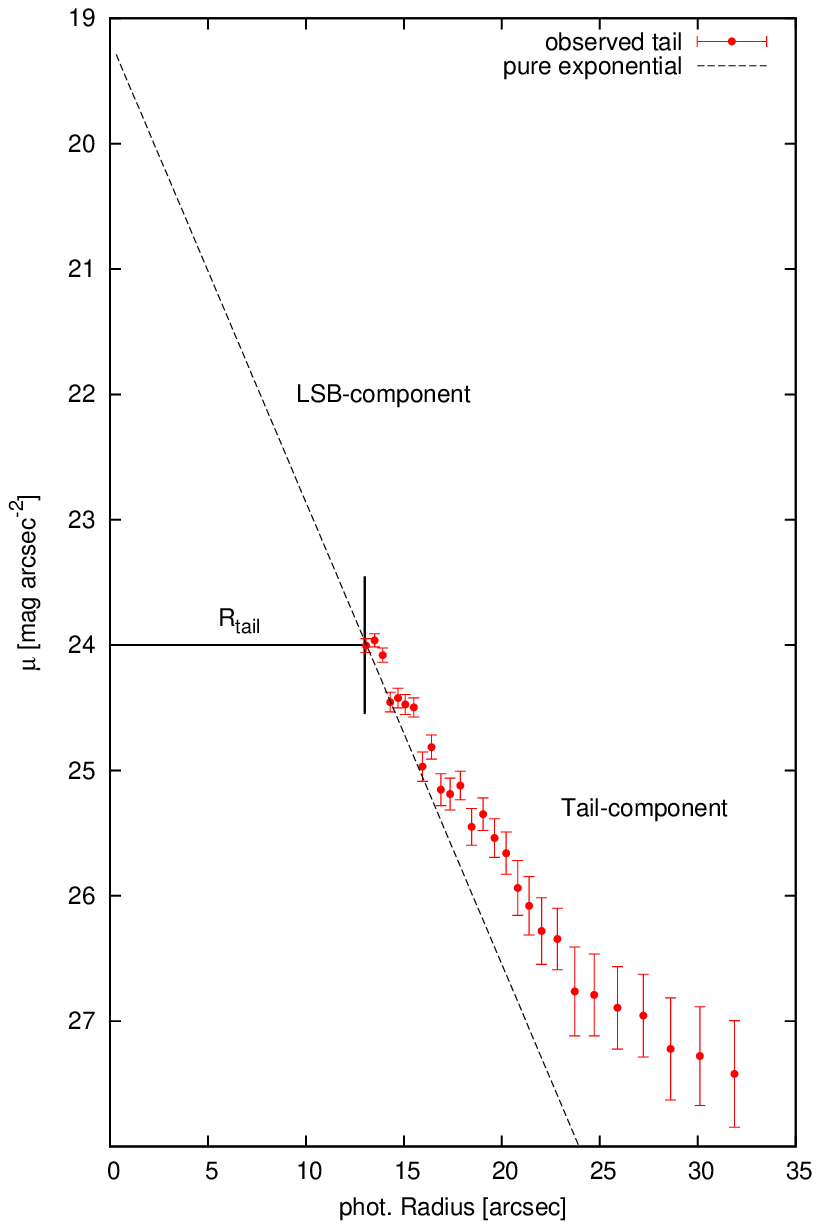}    
   \includegraphics[width=0.4\hsize]{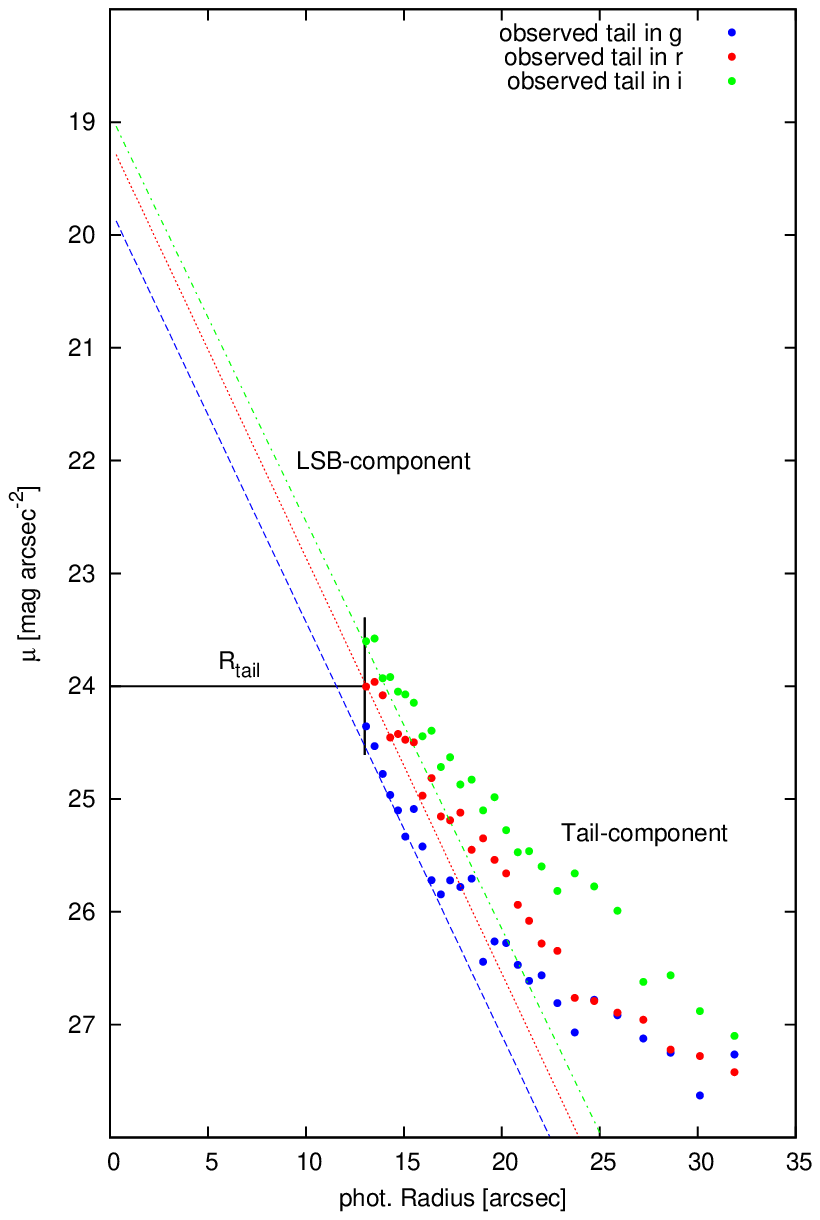}
         \caption{Surface brightness profile of a BCD {(VCC~0001)}. Left panel: The line corresponds to an exponential fit to the LSB-component and the red data points show the outer tail in the $r$-band. Right panel: Same, but for the g,r and i-filters. Obvious are the outer tails of the profiles at low surface brightness levels. The horizontal bars indicate the radius at which the profile {starts deviating} from a pure exponential fit.   
                       }
         \label{tail}
   \end{figure*}

The inspection of the surface brightness profiles of our BCD sample reveals that in {many} cases the {profile} of the LSB-component {flattens out at}
 large radii. This is illustrated in Fig.~\ref{tail} for VCC~0001, and is also clearly visible in Fig.~\ref{fig:examples} for VCC~1744 for radii $R^{\star}>15$\;arcsec. 
We note that such a flattening has been reported in the outskirts of some BCDs, e.g.\ II Zw 71 and 
Mrk 178, {where it was found to contribute no more} than $\sim$3\% of the total luminosity  \citet{Papaderos01b} and \citet{Papaderos02}.  {For some of our sample galaxies, however, already a quick estimate based on the surface brightness profile reveals a significant contribution (10\% for VCC~1744, see Fig.~\ref{fig:examples}). Since this leads to an increase in the derived effective radius, and since we will discuss in Section~\ref{sec:results} whether BCDs are smaller in size than early-type dwarfs, we cannot neglect this systematic effect on the radius.}

{
In the following, we will call the radius where the observed profile of the LSB-component starts deviating from a pure exponential slope \emph{tail radius} ${R_{\rm tail}}$ (Fig.~\ref{tail}). 
The exponentially fitted LSB-component was integrated up to  ${R_{\rm tail}}$ and was then continued with the observed data points of the tail. From this LSB+tail profile -- also taking the possible inner flattening into account -- we derive the Petrosian radius \citep{Petrosian76, Graham05a}, where the intensity drops to a fifth of the mean intensity within that radius \citep[][]{Blanton01a, Yasuda01}.}

LAZY, just like all surface photometry codes,
is very sensitive to the quality of the sky subtraction {and the objective selection and removal of fore/background sources. Especially in the faint extended outskirts of a galaxy, this can be difficult and} may introduce additional uncertainties.
To overcome this problem the parameters of the LSB-component were derived with LAZY {only for $R^{\star}\leq{R_{\rm tail}}$, and were} 
subsequently corrected for the missing flux. The latter was estimated by integrating the SDSS images
{
from ${R_{\rm tail}}$ out to two Petrosian radii, using elliptical apertures (see Section~\ref{sec:globalphoto}), masking contaminating sources, and correcting for the missing flux of these masked patches.
}
 This analysis resulted in a correction value $\delta {f}$ for the missing flux, such that the total flux {of the LSB-component} becomes
\begin{equation}
 	{ f_{\rm corrected, LSB} = f \left( \le R_{\rm tail} \right) + \delta f}\quad.
 \end{equation}

{
  The colours of the LSB-components were determined from the corrected total flux in the different bands.
  The effective radius ${R}_{\rm{eff,LSB}}$  was derived in the $r$-band by the integration of the combined LSB profile (inner flattening, exponential section, outer tail) to the radius within which half of the total flux is enclosed.
 The mean effective surface brightness $\left< \mu \right>_{\rm{eff,LSB}}$  was calculated as
\begin{equation}
 	{\left<\mu\right>_{\rm eff,LSB} = m_{\rm{LSB}} + 2.5 \log \left( 2 \pi  \, R_{\rm{eff,LSB}}^2 \right)}.
\end{equation}
}

\subsection{LSB-component: error estimation}
 The errors in our determination of magnitude and effective radius of
 the LSB component were estimated from a set of artificial BCD images,
 which we created based on the parameter range of the observational
 sample. These images consist of LSB-components with various profile
 shapes and a superposed starburst component with various
 strengths. They were analysed in the same manner as the Virgo BCDs,
 allowing us to compare input and output parameters.
 Details are given in Appendix~\ref{appendix}.

 The analysis of the artificial images reveals statistical errors of
$0.10$\,mag in ${M_r}$, $6\%$ in ${R_{\rm eff}}$, and $0.16$\,mag/arcsec$^2$ for ${\left<\mu\right>_{\rm eff}}$.
For the fainter
 BCDs of our sample, the radii derived with our method are somewhat
 smaller (11\%) than the true values, and the magnitudes are
 systematically brighter ($0.15$\,mag). These add up to a
 systematically brighter effective surface brightness by
 $0.41$\,mag/arcsec$^2$ for the fainter BCDs.
In the following figures we indicate the statistical errors with representative error bars and the systematic offsets with arrows pointing from the center of these error bars to the true value.

\subsection{Decomposition examples}
\label{example}


We illustrate our decomposition with two examples in Fig.~\ref{fig:examples}. VCC~1744 (left) shows a very strong star-forming region, which is off-centred. On the other hand, the star-forming regions in VCC~0130 (right) spread over the entire galaxy, resulting in an irregular LSB-component. 
The SDSS spectrum of VCC~1744 shows very strong emission lines, as is typical for BCDs. 
VCC~0130 also shows emission lines and a rising blue continuum, but 
 these emission lines are not as strong as in VCC~1744. 

The transition radius of VCC~1744 -- where the colours are becoming roughly constant, or the slope of the colour profile changes -- is $R_{\rm{trans}} \approx 4 \; \rm{arcsec}$ (see the colour profile). At smaller radii \rr the  starburst region starts to dominate. By extrapolating the exponential slope of VCC~1744 to ${R^{\star}=0}$\;arcsec one obtains a central surface brightness of $\mu_{\rm{0,LSB}} = 21.6 \;\rm{mag/arcsec^2}$, two magnitudes fainter than the value when including the starburst region.

In contrast, the profile of VCC~0130 shows that a 
pure exponential approximation of the LSB-component would overestimate the central surface brightness of the LSB-component. Therefore, we assume a central flattening and obtain 
$\mu_{\rm{0,LSB}} = 22.1 \; \rm{mag/arcsec}^2$, which by construction (see the \textit{modexp} fitting function in Section~\ref{innerflattening}) is $1.75$ mag/arcsec$^2$ fainter than the unmodified profile.
Since the optical image of VCC~0130 does show several star-forming regions, we consider the application of an inner flattening for separating the LSB and starburst component to be justified, even though the galaxy's colour profile is rather flat.



\section{Results}
\label{sec:results}

\begin{sidewaystable*}
\caption{Derived structural parameters of the BCDs and their LSB-components.}\tabularnewline
\begin{tabular}{l c c c c c c c c c c c c c}
\hline\hline
\label{parameters} 
VCC & ms & LSB & Flattening & ${(g-i)}_{\rm{tot}}$ & ${(g-i)}_{\rm{LSB}}$ & ${M}_{\rm{tot,}r}$ & ${M}_{\rm{LSB,}r}$ & ${R}_{\rm{eff,tot,}r}[\rm{kpc}]$ & ${R}_{\rm{eff,LSB,}r} [\rm{kpc}]$ & $\left<\mu\right>_{\rm{eff,tot,}r}$ & $\left<\mu\right>_{\rm{eff,LSB,}r}$ & D [Mpc] & ${M}_{\rm{tot,}r}$ \\                    
$\left[ 1 \right]$	& $\left[ 2 \right]$  &	$\left[ 3 \right]$	     &	$\left[ 4 \right]$					&	$\left[ 5 \right]$					     &	$\left[ 6 \right]$	    & $\left[ 7 \right]$  & $\left[ 8 \right]$ & $\left[ 9 \right]$ & $\left[ 10 \right]$ & $\left[ 11 \right]$  & $\left[ 12 \right]$  & $\left[ 13 \right]$  & $\left[ 14 \right]$  \\\hline
0001 & 2 & nE & 0 & 0.881 & 0.873 & -16.32 & -16.27 & 0.38 & 0.40 & 20.16 & 20.30 & 32.0 & -17.75\\ 
0010 & 1 & iE? & 0 & 0.735 & 0.754 & -16.09 & -15.87 & 0.37 & 0.42 & 20.36 & 20.84 & 32.0 & -17.52\\ 
0022 & 1 & nE & 0 & 0.638 & 0.734 & -15.45 & -15.30 & 0.39 & 0.42 & 21.07 & 21.42 & 32.0 & -16.89\\ 
0024* & 2 & nE & 0 & 0.695 & 0.765 & -16.24 & -16.01 & 0.36 & 0.42 & 20.12 & 20.69 & 32.0 & -17.68 \\ 
0074 & 1 & nE & 1 & 0.736 & 0.765 & -15.23 & -14.71 & 0.59 & 0.77 & 22.22 & 23.29 & 17.0 & -15.30\\ 
0130 & 1 & iI & 1 & 0.417 & 0.400 & -14.65 & -14.24 & 0.36 & 0.44 & 21.73 & 22.56 & 17.0 & -14.72\\ 
0144 & 2 & iE & 0 & 0.149 & 0.249 & -16.51 & -15.97 & 0.25 & 0.34 & 19.08 & 20.26 & 32.0 & -17.94\\ 
0172 & 1 & iI & 0 & 0.489 & 0.743 & -16.82 & -16.30 & 0.84 & 1.07 & 21.39 & 22.42 & 32.0 & -18.25\\ 
0207 & 1 & iI & 0 & 0.271 & 0.215 & -14.57 & -14.19 & 0.21 & 0.26 & 20.67 & 21.50 & 32.0 & -16.01\\ 
0223 & 1 & nE? & 0 & 0.623 & 0.758 & -15.37 & -15.01 & 0.32 & 0.41 & 20.74 & 21.63 & 32.0 & -16.81\\ 
0274 & 1 & iI & 0 & 0.473 & 0.622 & -14.40 & -14.34 & 0.55 & 0.60 & 22.87 & 23.13 & 32.0 & -15.84\\ 
0324 & 1 & i? & 0 & 0.476 & 0.630 & -17.46 & -17.25 & 0.96 & 1.17 & 21.03 & 21.68 & 17.0 & -17.52\\ 
0334* & 2 & iE & 0 & 0.526 & 0.698 & -16.02 & -15.72 & 0.43 & 0.54 & 20.74 & 21.53 & 17.0 & -16.08 \\ 
0410 & 2 & iI,C & 0 & 0.294 & 0.460 & -14.42 & -14.31 & 0.30 & 0.31 & 21.53 & 21.70 & 17.0 & -14.48\\ 
0428 & 2 & iI & 0 & -0.049 & 0.164 & -14.41 & -14.32 & 0.42 & 0.44 & 22.30 & 22.47 & 17.0 & -14.47\\ 
0459 & 2 & iI & 0 & 0.489 & 0.636 & -16.75 & -16.20 & 0.53 & 0.67 & 20.47 & 21.51 & 17.0 & -16.81\\ 
0513 & 1 & iI & 0 & 0.823 & 0.770 & -16.48 & -16.51 & 0.49 & 0.56 & 20.55 & 20.83 & 17.0 & -16.54\\ 
0562 & 2 & iI & 0 & 0.316 & 0.399 & -15.50 & -15.45 & 0.64 & 0.53 & 22.12 & 21.76 & 17.0 & -15.56\\ 
0641 & 2 & iI & 1 & 0.329 & 0.406 & -15.51 & -15.03 & 0.48 & 0.60 & 21.49 & 22.44 & 23.0 & -16.23\\ 
0802 & 2 & iI,C & 0 & 0.630 & 0.726 & -14.99 & -14.96 & 0.68 & 0.68 & 22.76 & 22.80 & 17.0 & -15.05\\ 
0841* & 2 & iE & 0 & 0.761 & 0.853 & -16.46 & -16.19 & 0.91 & 1.14 & 21.92 & 22.67 & 17.0 & -16.53 \\ 
0890* & 1 & iE? & 0 & 0.531 & 0.819 & -15.07 & -14.81 & 0.31 & 0.39 & 20.98 & 21.72 & 23.0 & -15.79 \\ 
1141 & 2 & nE & 0 & 0.733 & 0.738 & -15.50 & -15.34 & 0.40 & 0.46 & 21.08 & 21.57 & 23.0 & -16.22\\ 
1313 & 2 & iI & 0 & -0.263 & 0.058 & -14.50 & -13.66 & 0.18 & 0.26 & 20.33 & 22.03 & 17.0 & -14.56\\ 
1411 & 2 & iE & 0 & 0.650 & 0.768 & -16.19 & -16.20 & 0.98 & 1.09 & 22.34 & 22.58 & 17.0 & -16.25 \\ 
1437* & 2 & nE & 0 & 0.715 & 0.801 & -16.74 & -16.40 & 0.37 & 0.50 & 19.69 & 20.70 & 17.0 & -16.80 \\ 
1459 & 1 & iE? & 0 & 0.762 & 0.820 & -15.22 & -15.21 & 0.65 & 0.72 & 22.44 & 22.67 & 17.0 & -15.28\\ 
1572 & 1 & iE & 0 & 0.769 & 0.914 & -15.79 & -15.57 & 0.91 & 1.06 & 22.58 & 23.14 & 17.0 & -15.86\\ 
1744 & 2 & iI,C & 0 & 0.204 & 0.598 & -14.80 & -14.59 & 0.49 & 0.60 & 22.23 & 22.87 & 17.0 & -14.87\\ 
2015 & 1 & nE? & 0 & 0.655 & 0.771 & -15.49 & -15.36 & 0.60 & 0.66 & 21.99 & 22.33 & 17.0 & -15.55\\ 
\end{tabular}
\tablefoot{[1]: VCC numbers marked with ``*'' were also discussed in \citet{Lisker06b}; [2]: membership (ms) of the BCDs regarding the VCC; [3]: classification of the LSB-component (see Tab.~\ref{BCDsubtypes}); [4]  flattening: 1 = inner flattening was applied;  [13]: GOLDMine distance; [14]: magnitude with GOLDMine distance.}
\end{sidewaystable*}

   \begin{figure*}
   \centering
   \includegraphics[width=1.05\hsize]{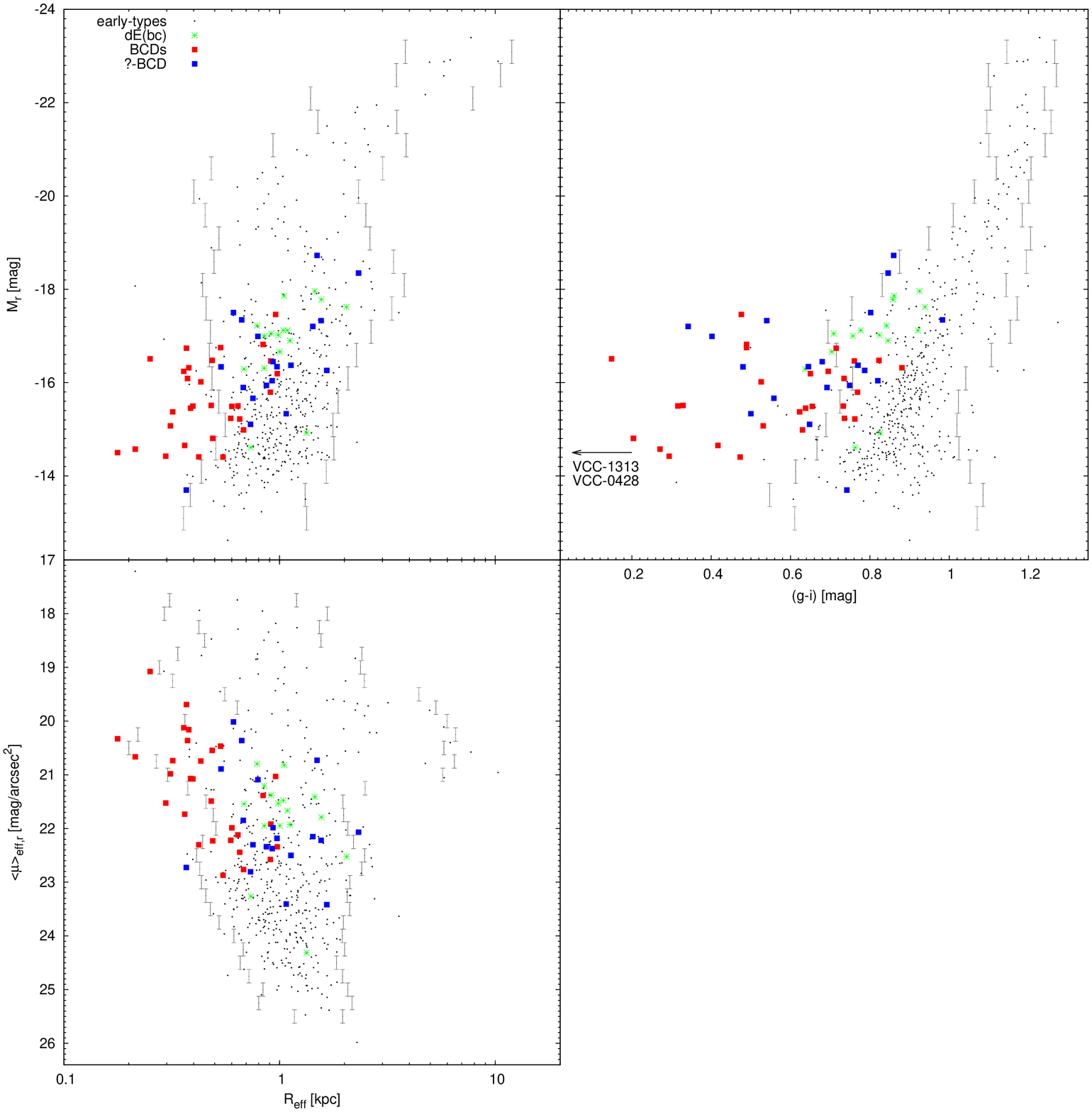} 
      \caption{The ${M_{r} - (g-i) - R_{\rm eff} - \left< \mu
          \right>_{\rm eff}}$-plane of BCDs and ETGs. No decomposition
        into starburst and LSB-component was applied. Red squares
        represent the primary BCD sample{, blue squares denote
          the candidate sample}. Black dots represent the sample of
        early-type galaxies taken from \citet{Janz08,Janz09}, for which colours were derived from the light within $2R_{\rm eff}$. Green asterisks are dE(bc)s from \citet{Lisker06b}. For each
        (vertical) interval in magnitude {or surface
          brightness}, the (horizontal) $\pm 2\sigma$-region of the
        ETGs  is indicated by the black bars. {VCC~0428 and 1313 have colours (${(g-i)}_{\rm{tot}} = -0.049$ and $-0.263 \; {\rm{mag}}$, respectively), lying outside the colour range of the plot and are therefore only indicated by an arrow.}
              }
         \label{all-values-BCDs}
   \end{figure*}

   \begin{figure*}
   \centering
   \includegraphics[width=0.49\hsize]{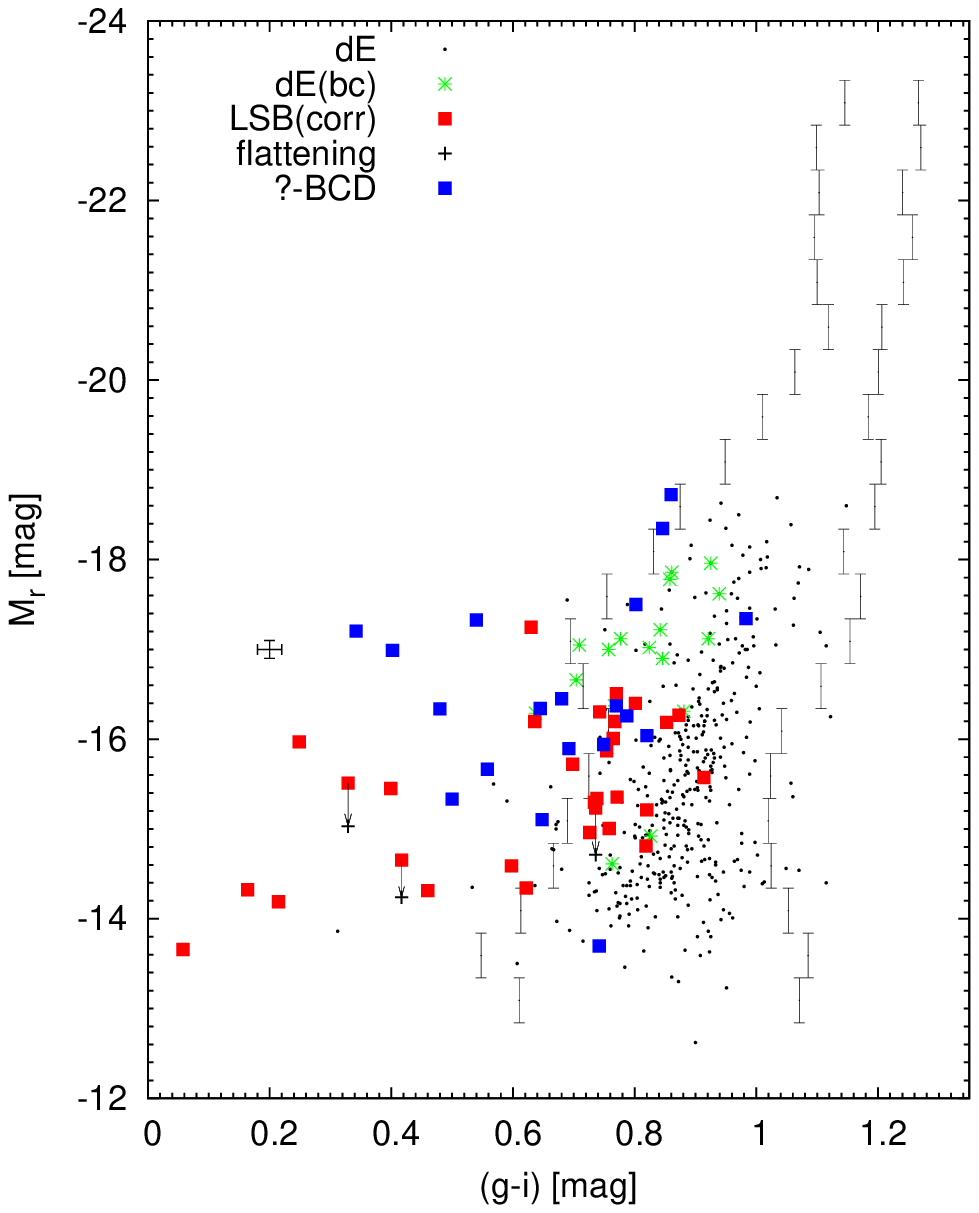} 
    \includegraphics[width=0.49\hsize]{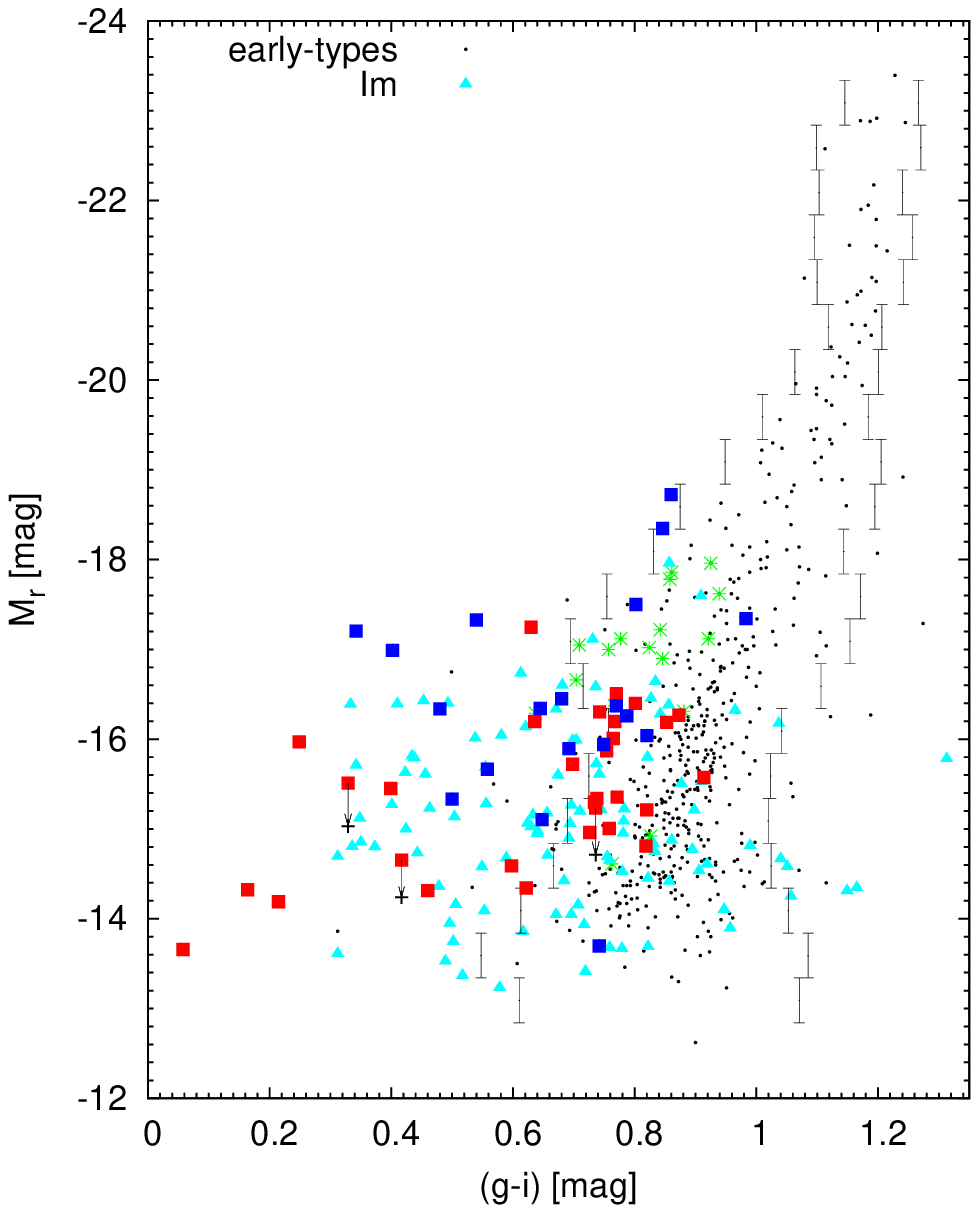} 
      \caption{{Left: Colour-magnitude diagram of the LSB-components of our primary BCD sample, compared to the Virgo early-type dwarf galaxies (labelled ``dE'' but including dS0s), for which colours were derived within $2R_{\rm eff}$. Red squares correspond to values of the LSB-components, where black crosses show LSB-components with inner profile flattening. The notation ``LSB(corr)'' indicates that the parameters of the LSB-components are corrected for the outer tail (see Section~\ref{tail-section}). The black vectors display the change of the parameters when an inner profile flattening is applied. Green asterisks correspond to dE(bc)s and blue squares to galaxies with a uncertain morphological classification (see section~\ref{mixedtypes}). Typical errors are shown on the left-hand side. Right: Same diagram, but additionally plotted are the irregulars (cyan triangles) and the entire population of early-type galaxies (black dots). In both panels we indicate the $2\sigma$-region of the entire ETG population with vertical bars. 
     }
              }
         \label{CMD}
   \end{figure*}

   \begin{figure*}
   \includegraphics[width=0.49 \hsize]{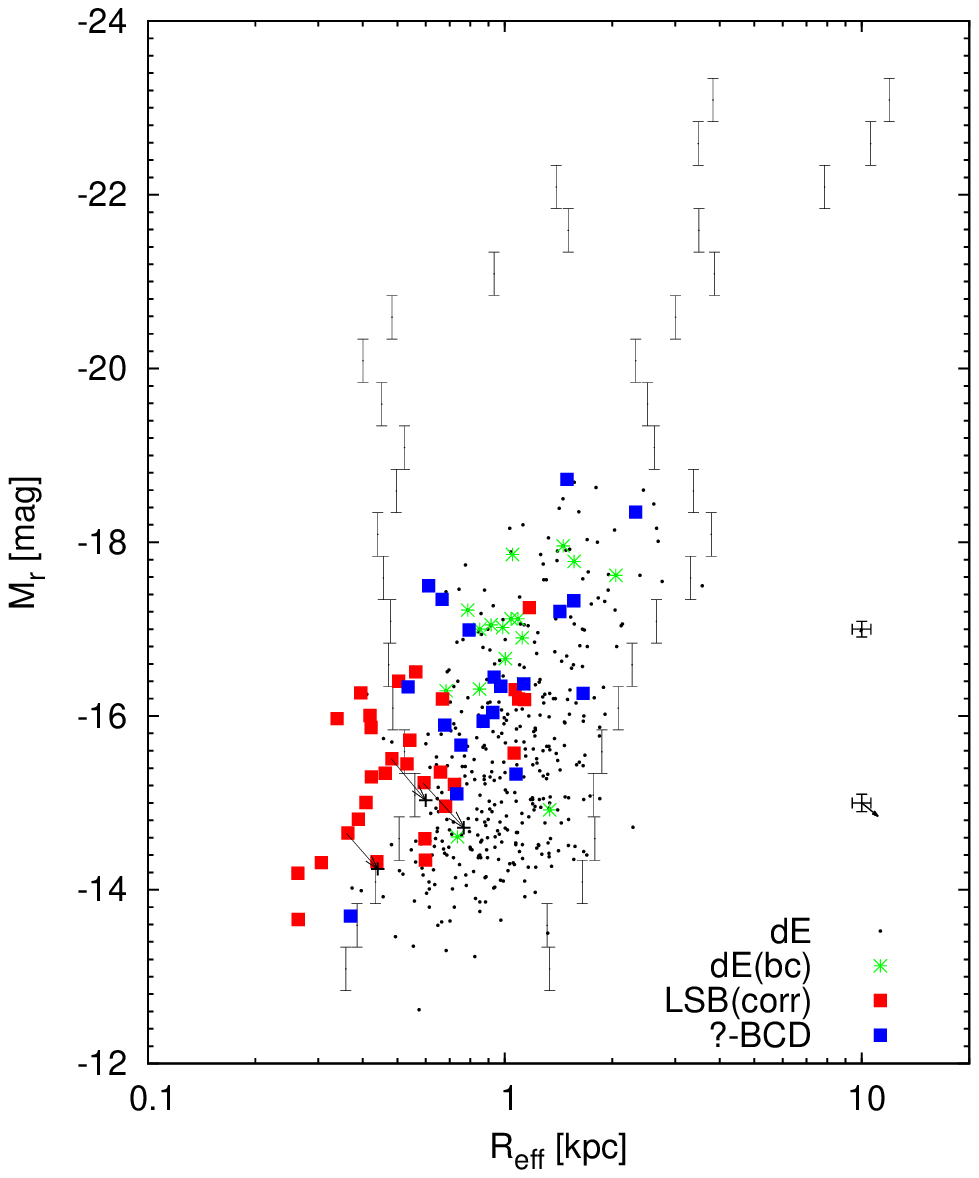} 
   \includegraphics[width=0.49 \hsize]{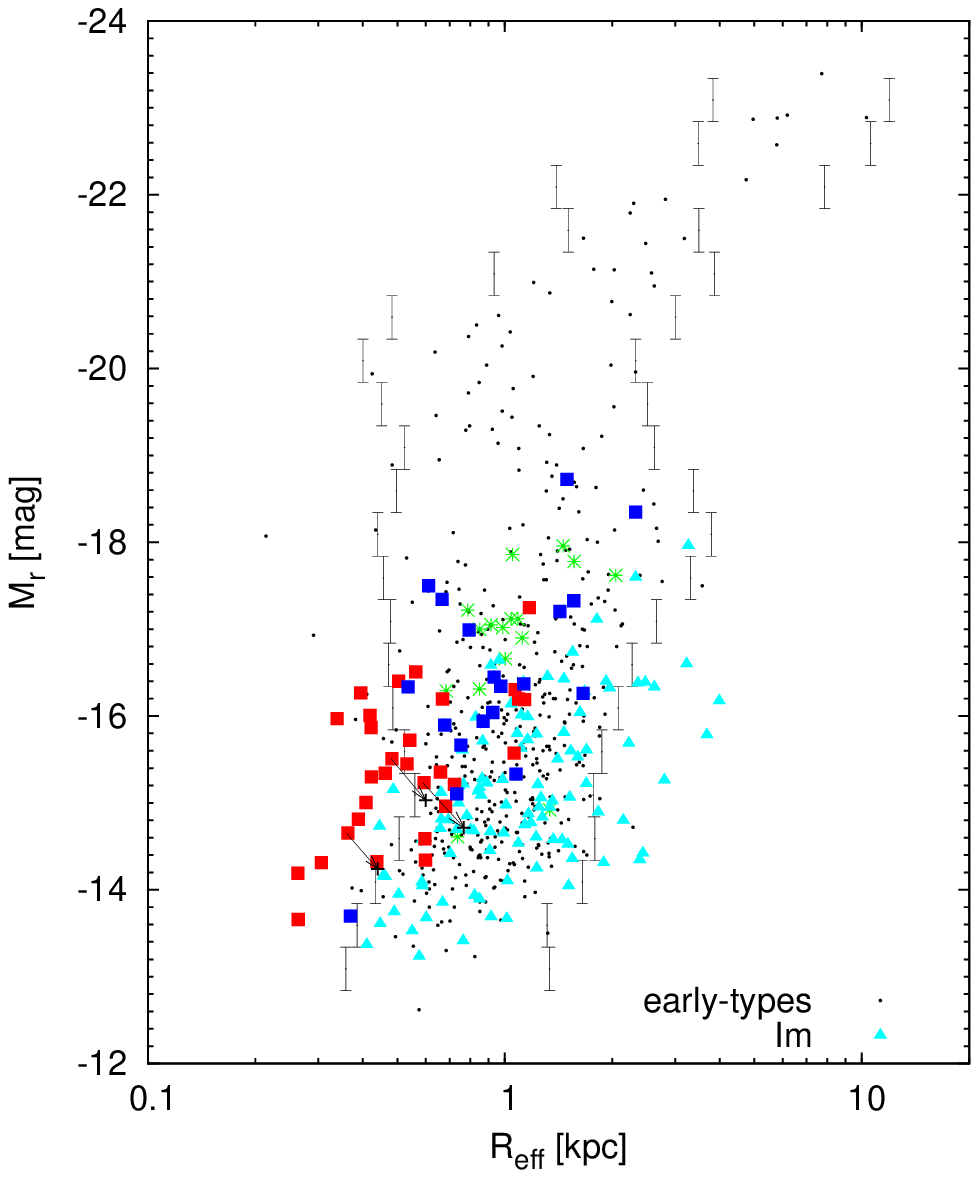}\\ 
   \includegraphics[width=0.49\hsize]{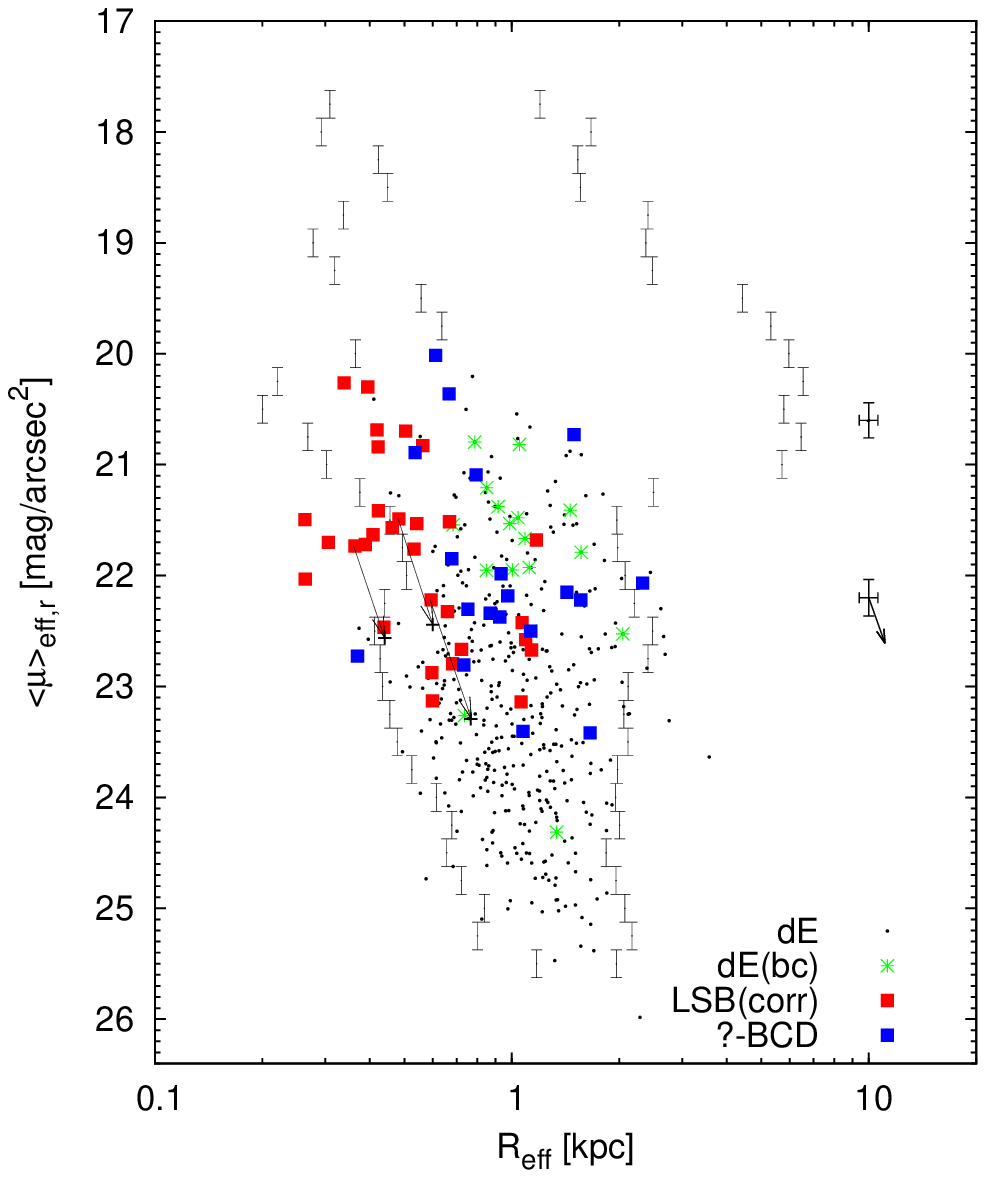} 
   \includegraphics[width=0.49\hsize]{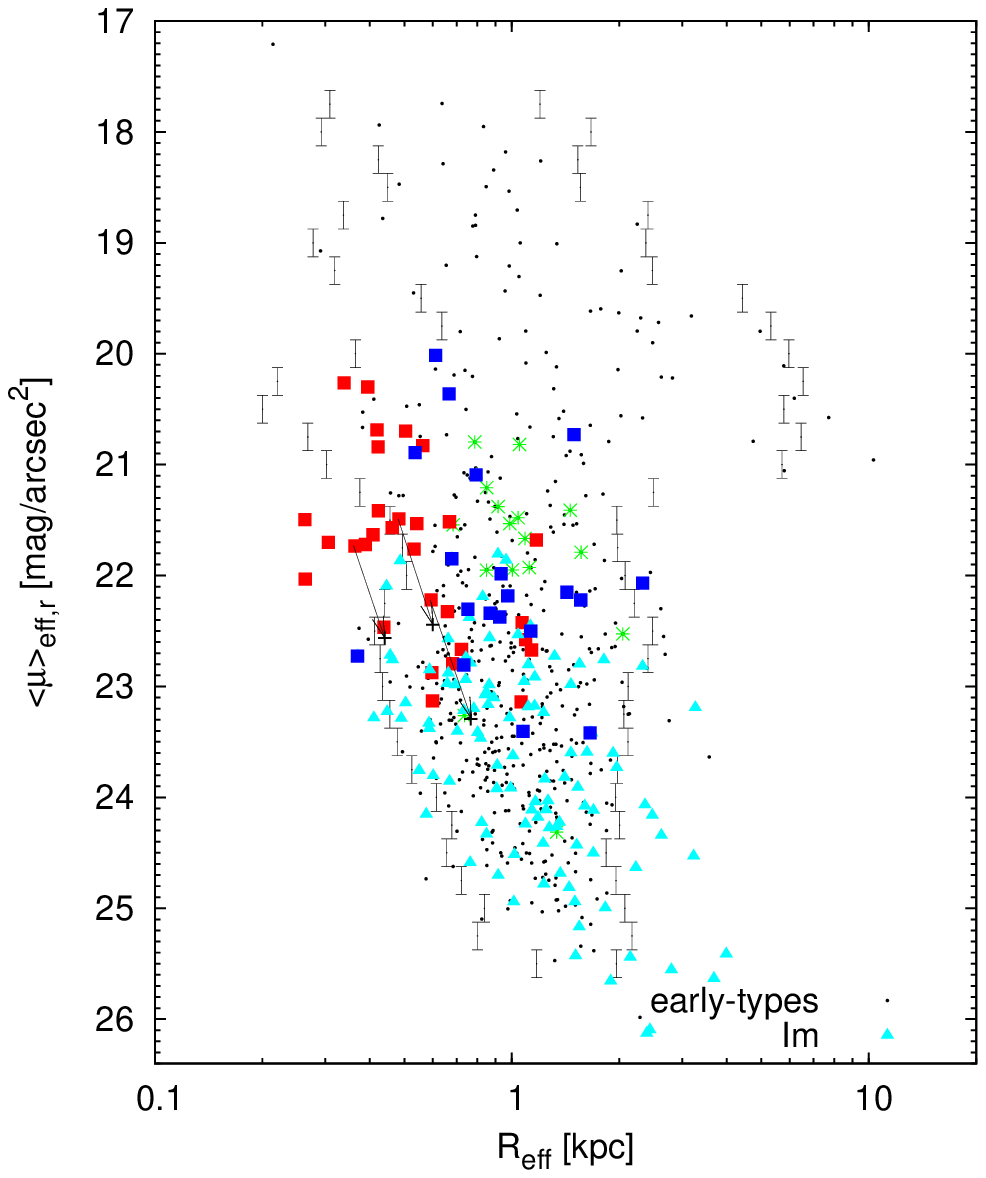}\\ 
      \caption{
{Effective radius ${R_{\rm eff}}$ vs.\ absolute magnitude (top panels) and mean effective surface brightness (bottom panels). Symbols are the same as in Fig.~\ref{CMD}. The black error bars in the left-hand panels illustrate the statistical errors, while the arrows indicate the systematic error, as derived from analysing artificial BCD images (see the Appendix). The error bars are placed on the ordinate so that they correspond to the average surface brightness and the magnitude of the respective set of artificial BCDs.  The arrows point in the direction towards which the measurements would have to be shifted to correct for the systematic error. Note that the upper arrow is almost invisible due to the very small error.}
              }
         \label{Mreff}
   \end{figure*}

   \begin{figure*}
     \centering
     \includegraphics[width=0.49\hsize]{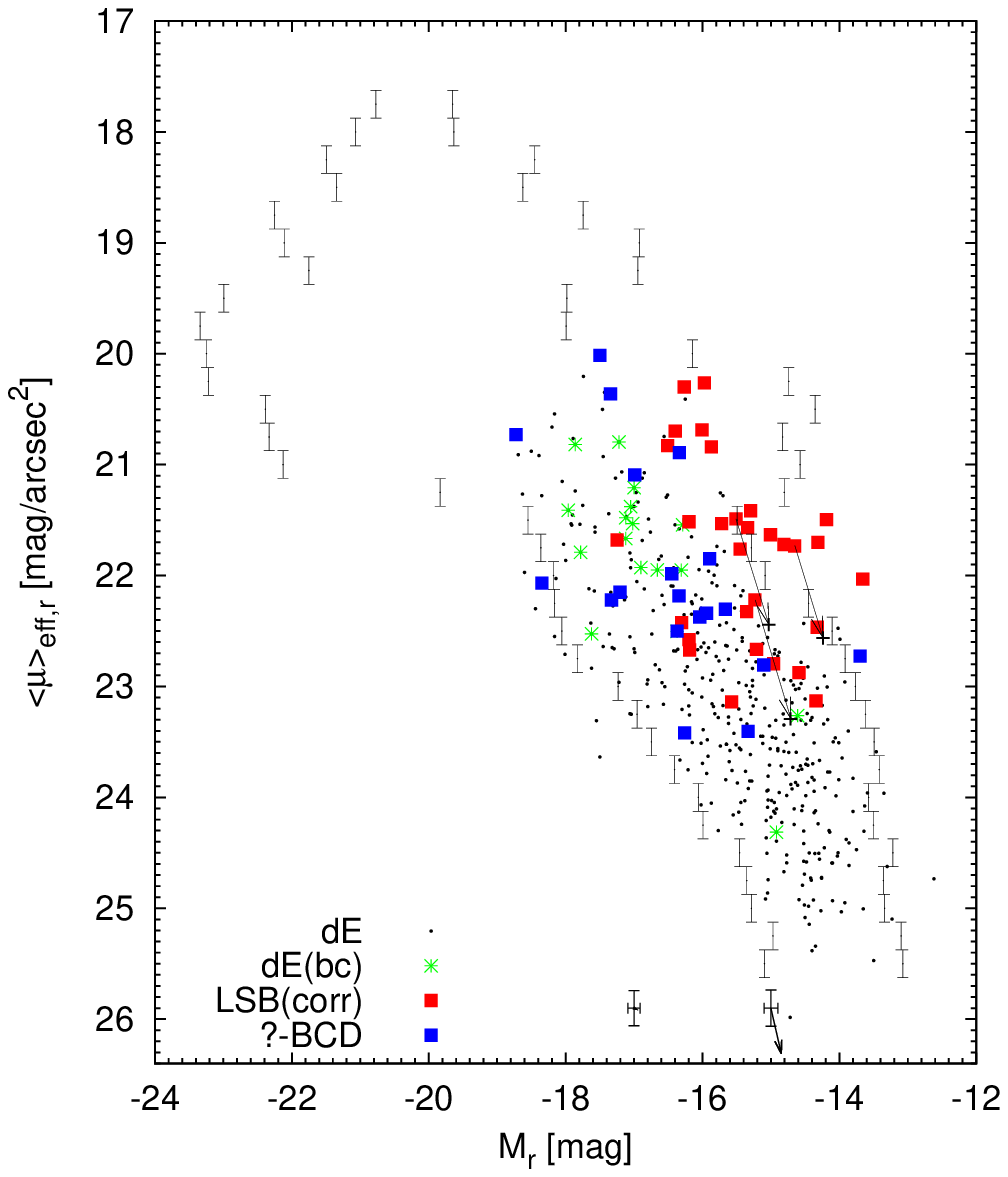} 
     \includegraphics[width=0.49\hsize]{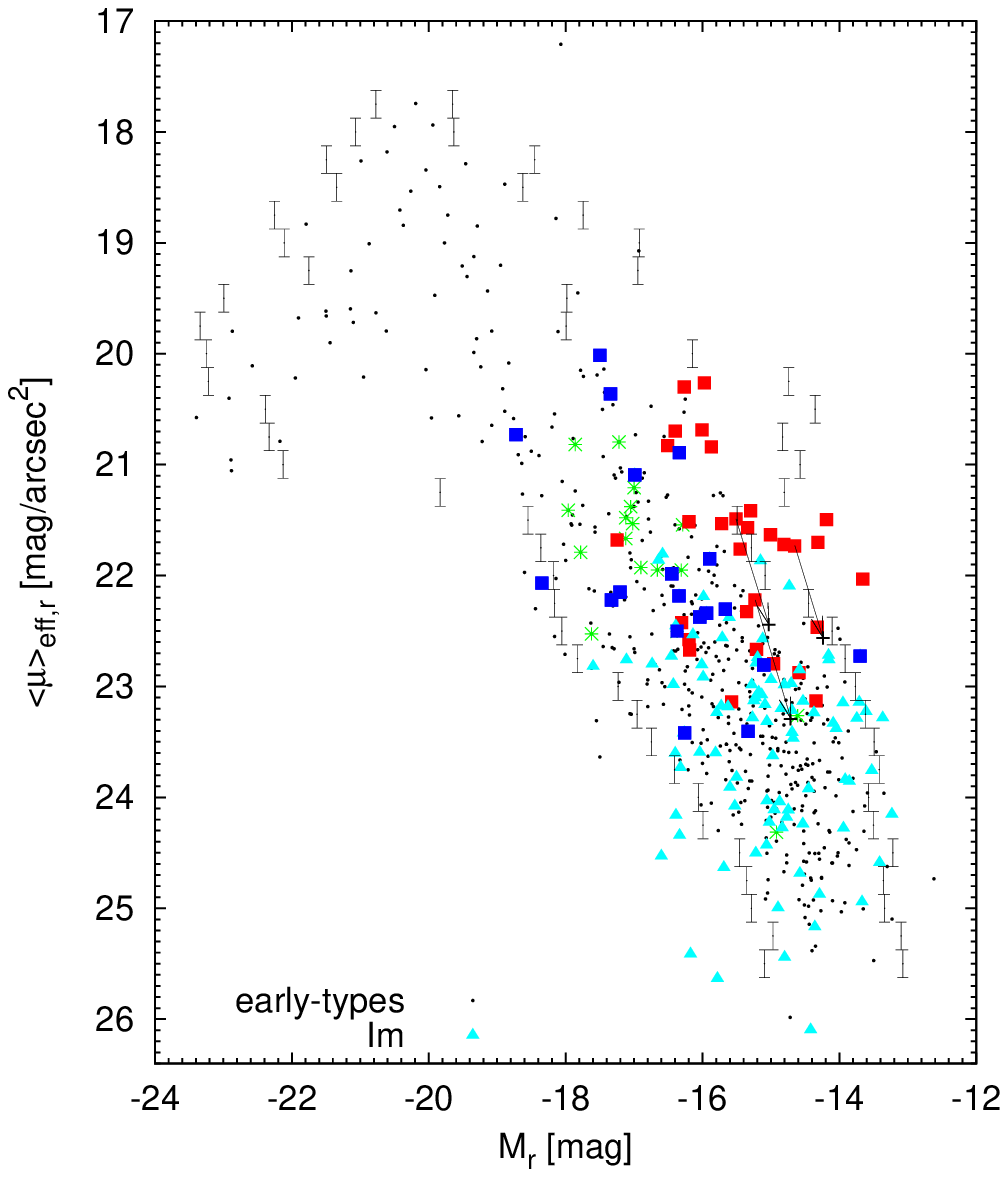} 
     \caption{Absolute magnitude ${M_r}$ vs.~mean effective surface brightness $\left<\mu\right>_{\rm{eff,r}}$ in r-band. Symbols are the same as in Fig.~\ref{CMD}.
{The black error bars in the left panel illustrate the statistical errors, while the arrows indicate the systematic error, as derived from analysing artificial BCD images (see the Appendix). The error bars are placed on the abscissa so that they correspond to the magnitude of the respective set of artificial BCDs.  The arrows point in the direction towards which the measurements would have to be shifted to correct for the systematic error. Note that the left arrow is almost invisible due to its very low value.} 
     }
     \label{Mr-mu}
   \end{figure*}

{The results of the photometric analysis are summarised in Tab.~\ref{parameters}. For each BCD values for the entire galaxy and for the underlying LSB-component are given} (column [5] to [12]). The membership (ms, column [2]) to the Virgo cluster was adapted  {from the VCC and later updates (see Section~\ref{sec:sample})}.
The BCD-classification according to Tab.~\ref{BCDsubtypes} is given in column [3].
Column [4] indicates whether a flattening towards smaller radii was applied (see Section~\ref{photometry} for more details).
Column [14] shows the total absolute magnitude of the entire BCD when
applying the distance {(column [13])} given by the GOLDMine database\footnote{http://goldmine.mib.infn.it/} \citep{Gavazzi03}. The differences between a distance of 16.5 Mpc and the GOLDMine distance are discussed in detail in Section~\ref{distance-issues}.

The structural and colour properties of the BCDs are illustrated in
Fig.~\ref{all-values-BCDs} for the entire galaxies, and in
Figs.~\ref{CMD} through \ref{Mr-mu} for only the LSB-components. {As a result of the decomposition, the LSB-components are less compact and redder than the entire galaxies.}

\subsection{Colour-magnitude diagrams}
\label{sectionCMD}
Figure~\ref{CMD} shows the colour-magnitude diagram (CMD)\footnote{This should not be confused with the colour-magnitude diagrams used in the studies of resolved stellar populations. We refrain from using the common term ``colour-magnitude relation'' (CMR) for Fig.~\ref{CMD}, as it may not be clear a priori whether an actual relation exists.} of the LSB-components (red squares) of our BCD sample. Galaxies with a profile flattening toward smaller radii are indicated with a black cross and the corresponding change in the magnitude is described by a vector. As mentioned in Section~\ref{innerflattening} the profile flattening parameters  are uncertain; therefore, the vector is to be regarded as an aid to the eye, pointing to the locus of the diagram where the true values are expected to be.

Blue squares show the results for galaxies with an uncertain morphological classification (see also Section~\ref{mixedtypes}).
Additionally, we plot with black dots  
 the early-type galaxies of the Virgo cluster \citep{Janz09}. According to the VCC these can be divided into the early-type dwarf (dE/dS0) and giant (E/S0) galaxies, based mainly on surface brightness and central concentration of light \citep{Sandage84, Binggeli85}. In {
Figs.~\ref{CMD} through \ref{Mr-mu}
}
 the early-type dwarfs are shown in the left panels, while all early types are shown in the right panels. The horizontal $2\sigma$-region of all early types is indicated in magnitude {or surface brightness} intervals by the vertical bars.  
The green asterisks correspond to {dE(bc)s, i.e.\ } early-type dwarfs with a blue core in the centre, taken from \citet{Lisker06b}.

On average the LSB-components of the BCDs are still bluer {than the} ETG population.
The LSB-components of the BCDs show a large 
 interval in colour with an average colour and {$2\sigma$-scatter} of ${\left<(g-i)\right>_{\rm LSB} = 0.64 \pm 0.45 \; mag}$, in contrast to the early-type dwarfs 
with  ${\left<g-i\right>_{\rm dE} = 0.88 \pm 0.19 \; mag}$.

There is {a very  blue BCD (VCC~1313) at ${(g-i)}_{\rm{LSB}} \approx 0.06 \; {\rm{mag}}$}. Such an extremely blue  ${(g-i)}_{\rm{LSB}}$-colour can naturally arise from extended nebular emission, as is the case for the XBCDs SBS~0335-052\,E \citep{Papaderos98a} and 
I~Zw~18 \citep{Papaderos02}. Extended nebular emission could diminish morphological asymmetries{; the structure
of VCC~1313 thus needs to be interpreted with some caution. In distant, poorly resolved starburst galaxies, extended emission can easily be mistaken for a stellar disc \citep{Papaderos02, Papaderos12}.}

{From Fig.~\ref{CMD} it can be seen} that the dE(bc)s are slightly offset from all early-type dwarfs on the CMD plot, showing 
bluer colours with an average value of $\left<{g-i}\right> = 0.81  \; {\rm{mag}}$. 
{The majority of them is brighter than the LSB-components of the BCDs.
While focusing on dE(bc)s, \citet{Lisker06b} also analysed  an additional sample of galaxies, which includes some BCDs of our study presented here (marked with ``*'' in Tab.~\ref{parameters})}. Due to the sample selection, all BCDs in the sample of \citet{Lisker06b} show regular elliptical isophotes, corresponding to the nE type.

\begin{table*}
\caption{{Average values of different ETG- and LSB-types.}}\tabularnewline
\label{averagecolours}
\centering
\begin{tabular}{c c c c c c c c c} 
\hline\hline 
Subtype & $\left<{g-i}\right>$ & $\sigma$ & $\left<{M_r}\right>$ &
$\sigma$ & $ \left<{R}_{\rm{eff,r}}\right>$ [kpc] & $\sigma$ [kpc] & $\left<\left<\mu\right>_{\rm{eff,r}}\right>$  & $\sigma$ \\ \hline
ETG & 0.944 & 0.702 & -16.252 & 2.042 & 1.192 & 0.831 & 22.501 & 1.844\\ 
dE & 0.887 & 0.106 & -15.732 & 1.295 & 1.079 & 0.432 & 22.883 & 1.286\\ 
dE(bc) & 0.814 & 0.084 & -16.812 & 0.925 & 1.096 & 0.343 & 21.885 & 0.878\\ 
BCDs (total) & 0.535 & 0.243 & -15.684 & 0.816 & 0.534 & 0.215 & 21.357 & 0.978\\ 
BCDs (LSB) & 0.647 & 0.217 & -15.488 & 0.796 & 0.613 & 0.264 & 21.841 & 0.900\\ 
nE (LSB) & 0.760 & 0.051 & -15.637 & 0.548 & 0.545 & 0.144 & 21.562 & 0.787\\ 
iE (LSB) & 0.711 & 0.213 & -15.806 & 0.490 & 0.684 & 0.331 & 21.701 & 1.165\\ 
iI (LSB) & 0.521 & 0.264 & -15.142 & 0.892 & 0.598 & 0.258 & 22.127 & 0.792\\
Im	& 0.694	& 0.217	& -15.166 & 0.977 & 1.239 & 0.695 & 23.598 & 0.999\\
\hline
\end{tabular}
\tablefoot{ The averages were determined for colours of ${(g-i) > 0 \; mag}$ to avoid extremely blue BCDs, which could be contaminated by strong nebular emission. $\sigma$ corresponds to the standard deviation of the mean. {The label ``dE'' refers to all early-type dwarfs. For the division of the ETGs into giants and dwarfs the VCC was used.}}
\end{table*}

The right hand side of Fig.~\ref{CMD} shows the same CMD but additionally the Virgo irregulars are plotted. No obvious separation between BCDs or irregulars can be found, since they {cover nearly the same region in the CMD, also shown by their average colours (Tab.~\ref{averagecolours}; see also \citealt{Roediger11a}).}  However, the irregulars tend to have redder colours at fainter magnitudes, with colours {comparable to the early-type dwarfs.
For the different subtypes of BCDs, the average $g-i$ colours show a tendency for BCDs with regular LSB-components (nE or iE) to be redder than irregular shaped LSB-components (see Tab.~\ref{averagecolours}).}

\subsection{Sizes of BCDs}
\label{BCD-size}

Figure~\ref{Mreff} shows the effective radius ${R}_{\rm{eff}}$ vs.\ the absolute magnitude  {(top panels)} and the mean effective r-band surface brightness $\left<\mu\right>_{\rm{eff,r}}$ {(bottom panels)} of the LSB-components and the ETGs. 
The LSB-components are small in size when compared to the total early-type dwarf population, but partly cover 
the same region as the compact $\sim$half of it, and mostly lie within the $2\sigma$-region of all early types.
{Towards brighter} $\left<\mu\right>_{\rm{eff,r}}$ the LSB-components become more compact. 
However, compensating for the systematic errors of the derived parameters would move the fainter LSB-components slightly further towards the early-type dwarf region (black arrow centered on the error bars in the left-hand panels of Fig.~\ref{Mreff}).

Table~\ref{averagecolours} summarises the average values of the different subtypes of dwarf galaxies{, showing that there are no significant morphological differences between the BCD subtypes.}
 Since nE-BCDs have similarities to the dE(bc)s it could have been expected that their sizes are comparable. 
However, we find that the nE-BCDs{, as well as the other BCD subtypes, are more compact than the dE(bc)s}. The difference would be even slightly larger,
if the (central) star-forming component  of the dE(bc)s was  also excluded for the derivation of their photometric parameters. The effect of this exclusion would, however, be much smaller in the dE(bc)s, since their central star-forming regions contribute less significantly to the total light of the galaxies than is the case for the BCDs.
{Comparing the LSB-components and the dE(bc)s
 in the
 ${M_{r} - R_{\rm eff}}$-plane (Fig.~\ref{Mreff}),
 one may conclude that dE(bc)s {fall into the linear extension} of the LSB-components toward larger ${R}_{\rm{eff}}$ and brighter magnitudes.}

The right hand side of Fig.~\ref{Mreff} shows additionally the irregular galaxies. In both diagrams one finds a 
 clear offset of the overall location of LSB-components and irregulars, 
with a transition region where both types can be found.
At a given ${R}_{\rm{eff}}$ the LSB-components of BCDs tend to be brighter and have a higher $\left<\mu\right>_{\rm{eff,r}}$. 

Figure~\ref{Mr-mu} shows the mean surface brightness ${ \left< \mu \right>_{\rm eff,r}  }$ as a function of the absolute magnitude ${M_r}$ for the LSB-components {and early-type dwarfs} on the left-hand side and {additionally for the irregulars and all early types on the right-hand side. The BCDs and irregulars show a different average location in this diagram,} but both overlap with the early-type dwarfs. {Still, several LSB-components are outside the $2\sigma$-region} of the early-type dwarfs.

%
\subsection{Apparent axis ratio of the LSB-component}
   
   \begin{figure}[ht!]
   \centering
   \includegraphics[width=\hsize]{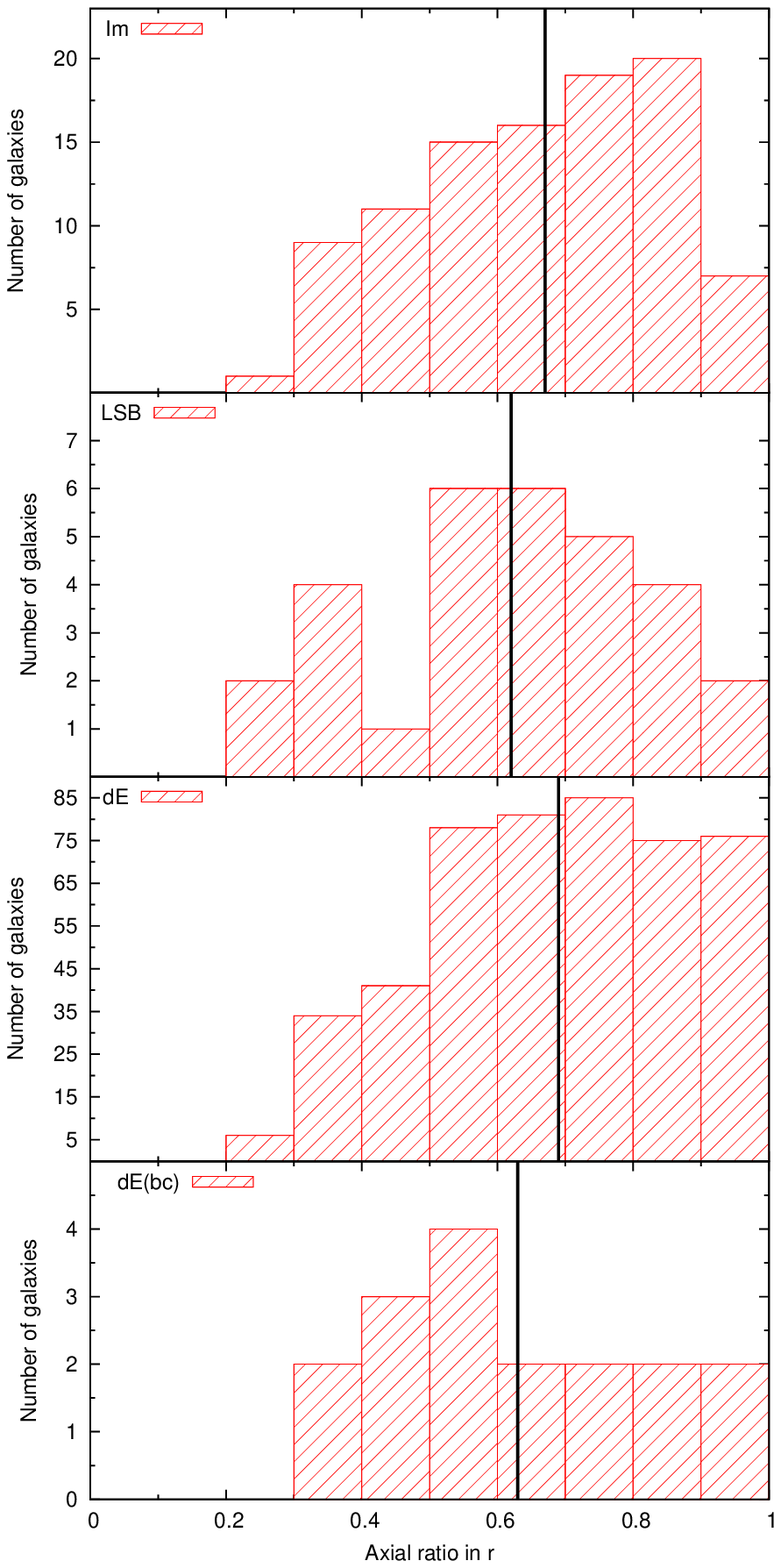} 
      \caption{Distributions of the axis ratios of Virgo cluster dwarf galaxies. The upper panels show the {irregulars and the LSB-components of the BCDs}. The early-type dwarfs (labelled ``dE'') also include the dE(bc)s, which are additionally plotted in the lower panel.
{The black vertical lines indicate the mean values.}
              }
         \label{axis-histo}
   \end{figure}
The axis ratios of the LSB-components have a mean value and standard deviation 
of $\left<{b/a}\right>_{\rm{LSB}} = 0.62 \pm 0.18$, {compatible} with the mean value of BCDs from \citet{Sung98a} ($\left<{b/a}\right>_{\rm{BCD,Sung}} = 0.67$) and very similar to the dE(bc)s ($\left<{b/a}\right>_{\rm{dE(bc)}} = 0.63 \pm 0.20$). The mean value of the early-type dwarfs is slightly larger -- i.e.\ they are slightly rounder -- with $\left<{b/a}\right>_{\rm{dE}} = 0.69 \pm 0.19$. Similarly, the irregulars have $\left<{b/a}\right>_{\rm{dE}} = 0.67 \pm 0.18$.

{Figure~\ref{axis-histo} shows} the distribution of axis ratios of the LSB-components and additionally the distribution of irregulars, early-type dwarfs, and specifically dE(bc)s.  
While the distribution of the dE(bc)s comes closest to what would be expected from the intrinsic shape of a (thick) disc, i.e.\ an oblate flattened galaxy, none of the distributions shown differ significantly from each other.
{In addition to what is displayed in the figure, no noticeable trend with magnitude is found.}
 Thus, no evolutionary connections can be ruled out based solely on the overall shape of the galaxy types. 

%
\subsection{Galaxies with uncertain morphological classification}
\label{mixedtypes}

\begin{table*}
\caption{{Uncertain BCD candidates and suggested reclassification (column 3).}}\tabularnewline
\label{transtab} 
\centering
\begin{tabular}{l c c c c c c} 
\hline\hline
VCC	& 	Type	& new Type & ${M}_{\rm{tot,r}}$ & ${R}_{\rm{eff,r}} [\rm{kpc}]$ & $\left<\mu\right>_{\rm{eff,r}}$	& ${(g-i)}$ 	\\\hline
0135 & Spec / BCD & - & -17.34 & 0.67 & 20.36 & 0.98 \\ 
0213 & dS? / BCD? & - & -17.50 & 0.61 & 20.01 & 0.80 \\ 
0281* & dS0 or BCD & - & -16.37 & 1.13 & 22.50 & 0.77\\ 
0309 & Im / BCD & Im & -15.33 & 1.07 & 23.41 & 0.50 \\ 
0429 & ? & Im / BCD & -15.10 & 0.73 & 22.81 & 0.65 \\ 
0446 & Im / BCD: & - & -15.94 & 0.87 & 22.34 & 0.75 \\ 
0655 & Spec,N: / BCD & - & -18.73 & 1.49 & 20.73 & 0.86 \\ 
1179 & ImIII / BCD & - & -16.34 & 0.97 & 22.18 & 0.64 \\ 
1356 & SmIII / BCD & BCD & -16.34 & 0.54 & 20.89 & 0.48 \\ 
1374 & ImIII / BCD & - & -16.99 & 0.79 & 21.09 & 0.40 \\ 
1427 & Im / BCD: & - & -16.45 & 0.93 & 21.98 & 0.68 \\ 
1713 & ? & Im / BCD & -15.66 & 0.75 & 22.30 & 0.56 \\ 
1725 & SmIII / BCD & - & -17.33 & 1.56 & 22.22 & 0.54 \\ 
1791 & SBmIII / BCD & - & -17.20 & 1.43 & 22.15 & 0.34 \\ 
1804 & ImIII / BCD & - & -16.04 & 0.92 & 22.37 & 0.82 \\ 
1955 & Spec / BCD & dE(bc)? & -18.35 & 2.33 & 22.07 & 0.85 \\ 
1960 & ImIII / BCD? & - & -13.70 & 0.37 & 22.73 & 0.74 \\ 
2007 & ImIII / BCD: & - & -15.89 & 0.68 & 21.85 & 0.69 \\ 
2037 & ImIII / BCD & Im & -16.26 & 1.66 & 23.42 & 0.79\\
\hline
\end{tabular}
\tablefoot{* from \citet{Janz09}}
\end{table*}

{As outlined in Section~\ref{sec:sample}, in addition to our primary working sample there are 21 uncertain BCD candidates in the VCC, 19 of which are covered by the SDSS DR5 (Tab.~\ref{transtab}).}
Using the  parameter space (${M_r-(g-i)_{\rm LSB}-R_{\rm eff}-\left<\mu\right>_{\rm eff}}$) of the BCDs {described in the previous sections, as well as the visual inspection of images}, we are able to {assess} whether these galaxies {should be reclassified as genuine BCDs or not}.

VCC~1374, classified as ``Im/BCD'', falls within the $R_{\rm eff}$ and $\left<\mu\right>_{\rm eff}$ values of the primary BCDs, but is brighter than nearly all of them. It is also not far from the bright end of the irregular galaxies' locus. However, its appearance is indeed that of an irregular or Sm-type galaxy with strong star-forming regions, hence appropriately described by ``Im/BCD''. The relatively small effective radius of $0.79$\,kpc is merely a result of its very elongated shape (axis ratio $0.32$) implying an edge-on view --- the half-light semimajor axis is $1.39$\,kpc instead.

For the galaxies VCC~0309 and VCC~2037, {also classified as ``Im/BCD''}, we found that {their} parameters are in the same region where the irregular galaxies 
are located{, which is also consistent with their visual appearance. We thus suggest to reclassify both as ``Im''.}
For the rest of the galaxies {classified as ``Im/BCD'',} a clear distinction between BCDs and irregulars is not possible, because they 
are located in the overlap zone between these two {types}. Their class thus remains unchanged.

There are three galaxies initially classified as ``Sm / BCD'' and ``SBm / BCD'', respectively. 
VCC~1356 is located in the same parameter {region} as the BCDs{, consistent with its visual appearance. We thus suggest to reclassify it to ``BCD''.} VCC~1725 and VCC~1791 do not show a clear alikeness 
or separation from the BCDs{; their class remains unchanged.}

The galaxy VCC~0281 was taken from {the early-type sample of \citet{Janz09} and was classified in the VCC as ``dS0 or BCD''.} 
It was also described by \citet{Lisker06b} as a dE(bc).
In the ${(g-i)-M_r}$-plane, VCC~0281  {falls into the same region as several of the brighter BCDs. However, it lies at the edge of the structural parameter range of the BCDs, and well inside the early-type dwarf regime. We thus refrain from a reclassification.}

\begin{figure}[ht!]
\centering
\includegraphics[width=\hsize]{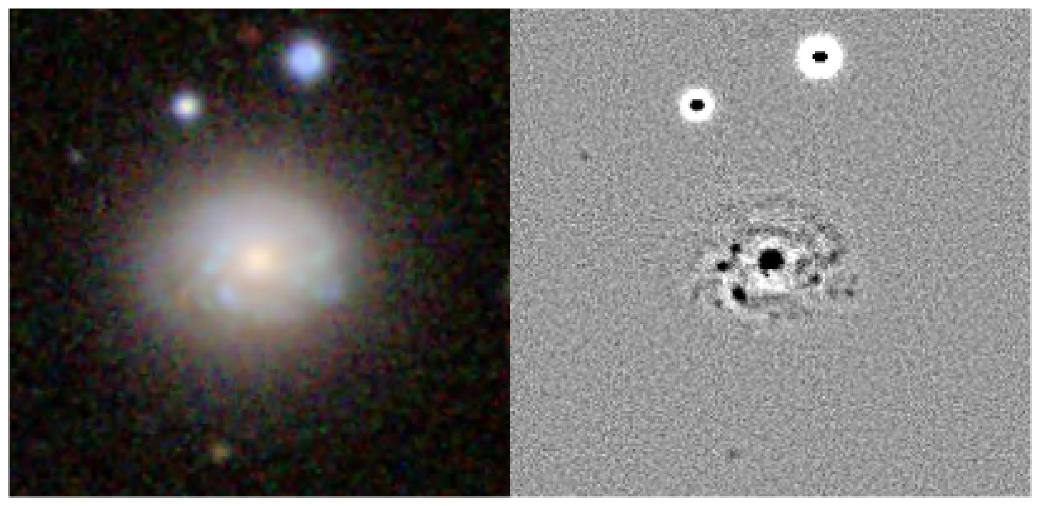}
\caption{{Image (left) and unsharp mask (right) of galaxy VCC~0213. Both panels have a width of $80"$, corresponding to $6.3$\,kpc at a distance of $16.5$\,Mpc. The unsharp mask was created using a two pixel Gaussian kernel.}
}
\label{sdss-vcc0213}
\end{figure}

The classification of VCC~0135, 0655, and 1955 (``Spec/BCD''), as well as VCC~0213 (``dS?/BCD?''), implies the presence of hints for spiral structure.
Indeed, VCC~0213 shows spiral structure in the SDSS image, and also for VCC~0135, a  weak spiral-like feature is visible. For VCC~0655, an inner ring or spiral-like structure is mimicked by several star-forming knots. No spiral structure can be seen for VCC~1955. 
In the classification scheme of \citet{Sung02}, they reported {that the ``Postmerger'' BCDs of their sample sometimes showed ``spiral structures with a compact off-centred core''}. 
However,
to the knowledge of the authors, there is no detailed study about BCDs with spiral structure visible at optical wavelengths. HI observations of BCDs by \citet{vanZee01} revealed rotation in BCDs, but the optical morphology of the used sample has smooth, symmetrical isophotes.

VCC~0213 is located at the bright and red end of the BCD region.
In terms of structure, it has similar parameters as the BCDs.
Its red core (Fig.~\ref{sdss-vcc0213}) seems to contradict the classification as a BCD. The spiral structure, together with the red core -- which appears similar to a bulge -- might hint at a spiral galaxy with a greater distance than the one of the GOLDMine data base of 17.0 Mpc. However, the heliocentric velocity according to GOLDMine is ${v_{\rm helio}= -165 \; km/s}$, which is consistent with being a Virgo member (cf.\ Fig.~\ref{goldmine-velocity}) and makes it unlikely that the galaxy is a background object. We thus decide to leave its classification unaltered.

VCC~0135 lies at the edge of the structural locus of the brighter BCDs.
However, its (g-i)-colours are too red, and its image looks like that of a low-luminosity lenticular or early-type spiral galaxy. The spectrum shows strong emission lines in the center, where blue {features} can be seen upon detailed inspection of the ``core-region''. Therefore, we refrain from changing its classification.

VCC~0655 seems to be too bright ($-18.73$\,mag) {and too large (${R}_{\rm{eff}}=1.49$\,kpc)
compared to the BCDs. However,} the image displays several star-forming regions in the central part of the galaxy, which are surrounded by a redder stellar host, typical for BCDs. Furthermore, VCC~0655 has some similarities to the well-known BCD NGC 2537 (Mrk 86, ``Bear's Paw Galaxy''). The sample of BCDs in \citet{Papaderos08} also shows a wide spread in radius from very compact objects up to objects with ${R_{\rm eff} \gtrapprox 1 \;  kpc}$. Thus, it cannot be ruled out that VCC~0655 is a bright BCD. We refrain, however, from reclassifying it, since deeper observations are needed to undoubtedly judge whether spiral structure is present.

At first glance, VCC~1955 appears similar to VCC~0655, except that its star-forming region is more centrally concentrated. However, its surface brightness is much lower, and in turn it has an even larger effective radius, too large compared to the other BCDs. Given its overall red colour and regular shape, VCC~1955 would qualify as a large-sized dE(bc), i.e.\ an early-type dwarf with a blue core. We therefore reclassify this object as ``dE(bc)?'', but remark that deeper observations might reveal further structural features.

Finally, as mentioned in Section~\ref{sec:sample}, visual inspection of VCC~0429 and VCC~1713 (classified as ``?'') suggests a reclassification to ``Im / BCD''. Indeed, their parameters fall in the overlap region of BCDs and irregulars. VCC~1411 was reclassified from ``pec'' to ``BCD?'' based on visual inspection.

The newly determined morphological types of the galaxies are {listed} in Tab.~\ref{transtab}. For one galaxy (VCC~1356) we suggest a reclassification to a primary BCD class. For three other objects (VCC~309, 1955, and 2037), a {unique} type could be assigned, instead of a mixed type. All other galaxies indeed show  complex properties that can only be described with a combination of two types.



\section{Discussion}

\begin{table}
	\caption{Comparison of the BCDs regarding their membership class.}\tabularnewline
	\label{diff-certain-poss}
	\centering
	\begin{tabular}{lcccc}
	\hline\hline
	Parameter				& Certain	&	$\sigma$ &  Possible	&	$\sigma$\\ \hline
$\left<M_r\right>$ [mag] & -15.68 & 0.84 	& -15.58 & 0.88 \\ 
$\left<g-i\right>$ [mag] & 0.44 & 0.32 	& 0.60 & 0.16 \\ 
$\left<R_{\rm{eff,r}}\right>$ [kpc] & 0.49 & 0.22 	& 0.54 & 0.24 \\ 
$\left<\mu\right>_{\rm{eff,r}}$ $\rm{[mag/arcsec^2]}$ & 21.15 & 1.10 	& 21.47 & 0.83 \\ 
	\hline                                   
	\end{tabular}
\tablefoot{{The average colour values are given for the entire
    galaxies without a decomposition. They were determined only from values with $(g-i)>0$ to avoid contamination by extreme outliers.}}
\end{table}

From Tab.~\ref{diff-certain-poss} it is apparent that various photometric parameters of {certain and possible Virgo cluster members agree within $1\sigma$}.
Therefore, in the following discussion no differentiation based on Virgo membership was applied.

\subsection{Distance dependence of the results}
\label{distance-issues}

\begin{table*}
\centering
\caption{GOLDMine distances of the different Virgo clouds. $\Delta {m}$ and $\alpha({R_{\rm eff}})$ indicate the {magnitude offset and the relative increase in radius} if the GOLDMine distances are used instead of a constant distance of 16.5 Mpc.}\tabularnewline
\label{tab:distances}
\begin{tabular}{c l c c}
\hline\hline
Distance			&	Location in Virgo cluster	&	$\Delta$m [mag] 		&	$\alpha({R_{\rm eff}})$	\\
 $\rm{[Mpc]}$		&						&	to d=16.5 Mpc			&	to d=16.5 Mpc			\\\hline
17				&	E-, N-, S-cloud, A-cluster	&	0.06					&	1.03					\\
23				&	B-cluster				&	0.72					&	1.39					\\
32				&	M-, W-cloud			&	1.44					&	1.94					\\
\hline                                   
\end{tabular}
\end{table*}

The GOLDMine data base provides {individual distances to} the BCDs, which can be found in column [13] in Tab.~\ref{parameters}.
{
The distances are based on their association with various substructures in the projected Virgo cluster area \citep{Gavazzi99}. These were defined based on the Tully-Fisher relation \citep{Tully77} for a sample of spiral galaxies and the Fundamental Plane \citep{Djorgovski87} for E/S0 galaxies, revealing separate regions in the parameter space of velocity and distance.
}
 Table~\ref{tab:distances} shows the different distances and the
 resulting {magnitude offsets when adopting the GOLDMine
   distances instead of $d = 16.5$ Mpc ($m-M=31.09$) mag}. The factor by which ${R_{\rm eff}}$ [kpc] changes due to different distances is given by $\alpha({R_{\rm eff}})$.

Since the allocation of the BCDs to the different substructures of the Virgo cluster only depends on the \emph{projected} location of the galaxies, the GOLDMine distances can not be taken for granted. For instance, the column depth of the different substructures {-- and of the main cluster itself -- cannot be taken into account.}

\begin{figure}[ht!]
\centering
\includegraphics[width=\hsize]{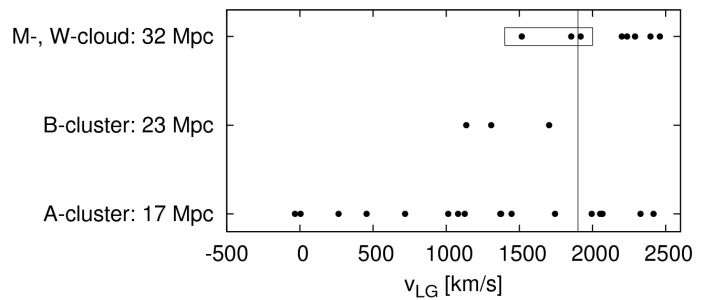} 
\caption{GOLDMine distance vs.~the velocity of the Virgo BCDs. The vertical solid line at ${v_{\rm LG}= 1900 \; km/s}$ corresponds to the limit from which galaxies belong to the M and W cloud {according to} \citet{Gavazzi99}. The {three BCDs} in the box {lie in the projected cloud region, but} are below or {only} slightly above this velocity limit. {We therefore adopt a distance of 16.5 Mpc for them instead of the  GOLDMine distance of 32 Mpc.}  
               }
\label{goldmine-velocity}
\end{figure}

\begin{figure*}[ht!]
\includegraphics[width=0.49\hsize]{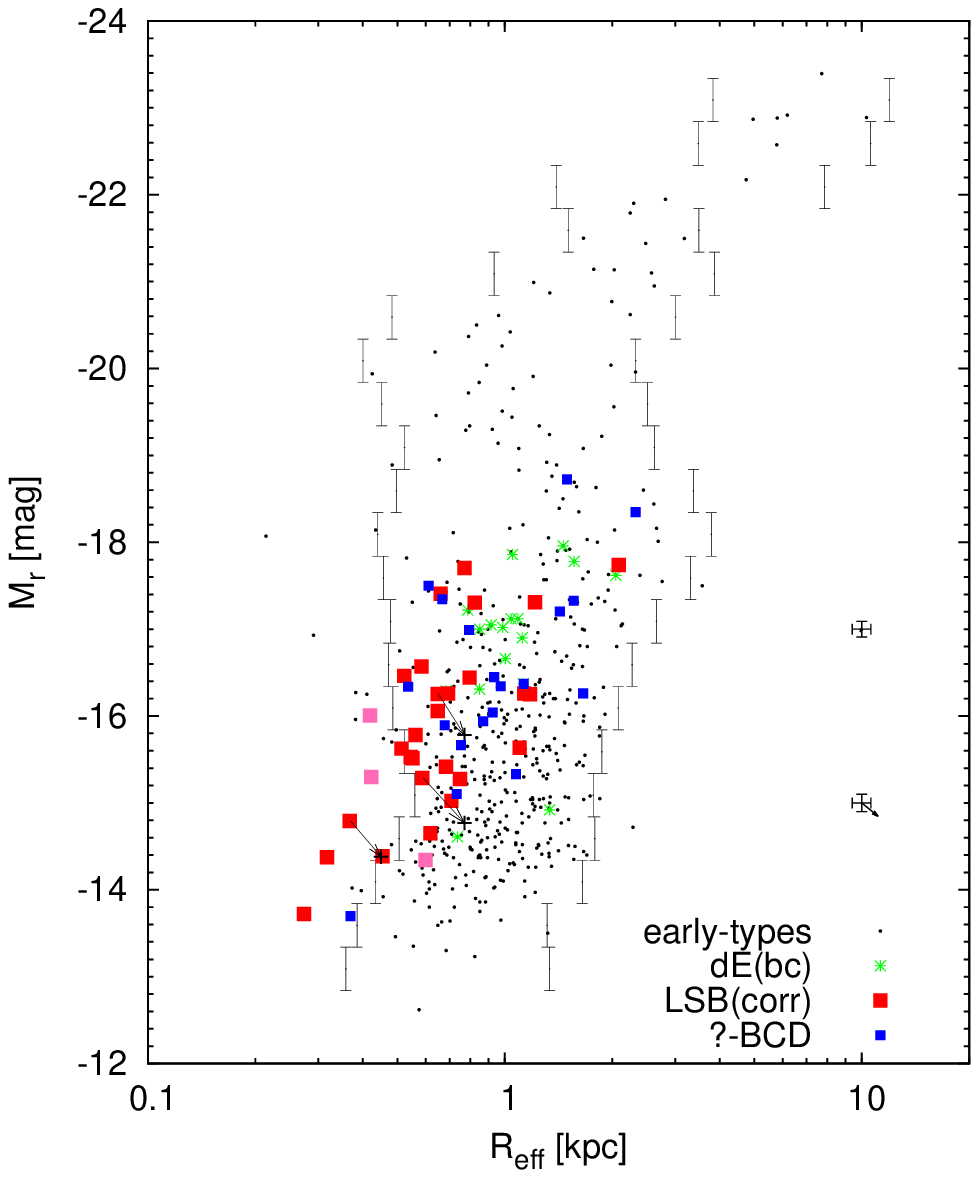} 
\includegraphics[width=0.49\hsize]{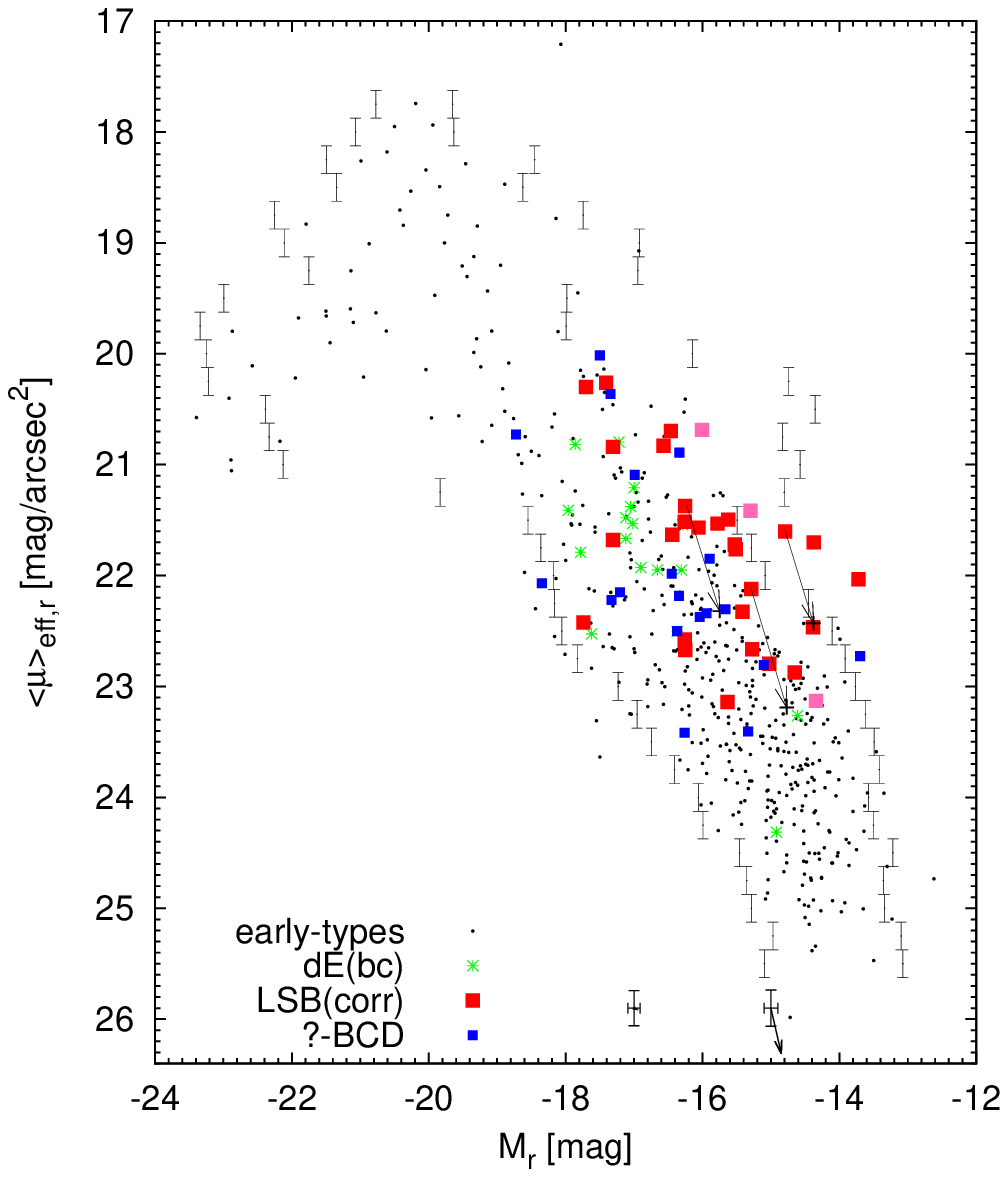}\\ 
\includegraphics[width=0.49\hsize]{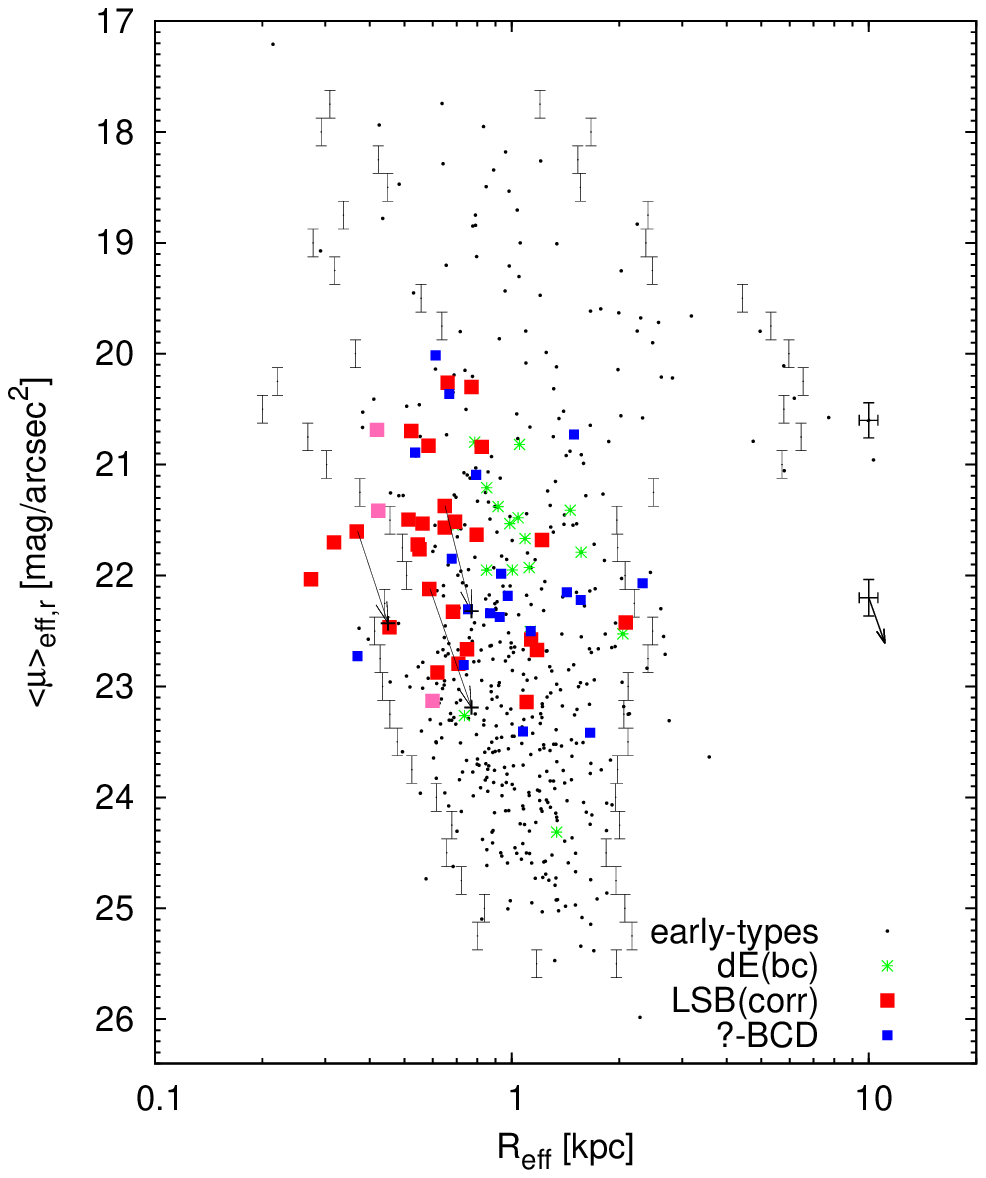}\\ 
\caption{{Same as Figs.~\ref{Mreff} and \ref{Mr-mu}, but instead of a constant distance of 16.5 Mpc, the distances given by the GOLDMine database are used for the LSB-components of the primary sample, except for the three objects (shown with magenta symbols) marked in Fig.~\ref{goldmine-velocity}, which remain at 16.5 Mpc. All other galaxy types also remain at a constant distance of 16.5 Mpc.
}
}
\label{goldmine-distance}
\end{figure*}

   \begin{figure*}[ht!]
    \includegraphics[width=0.49\hsize]{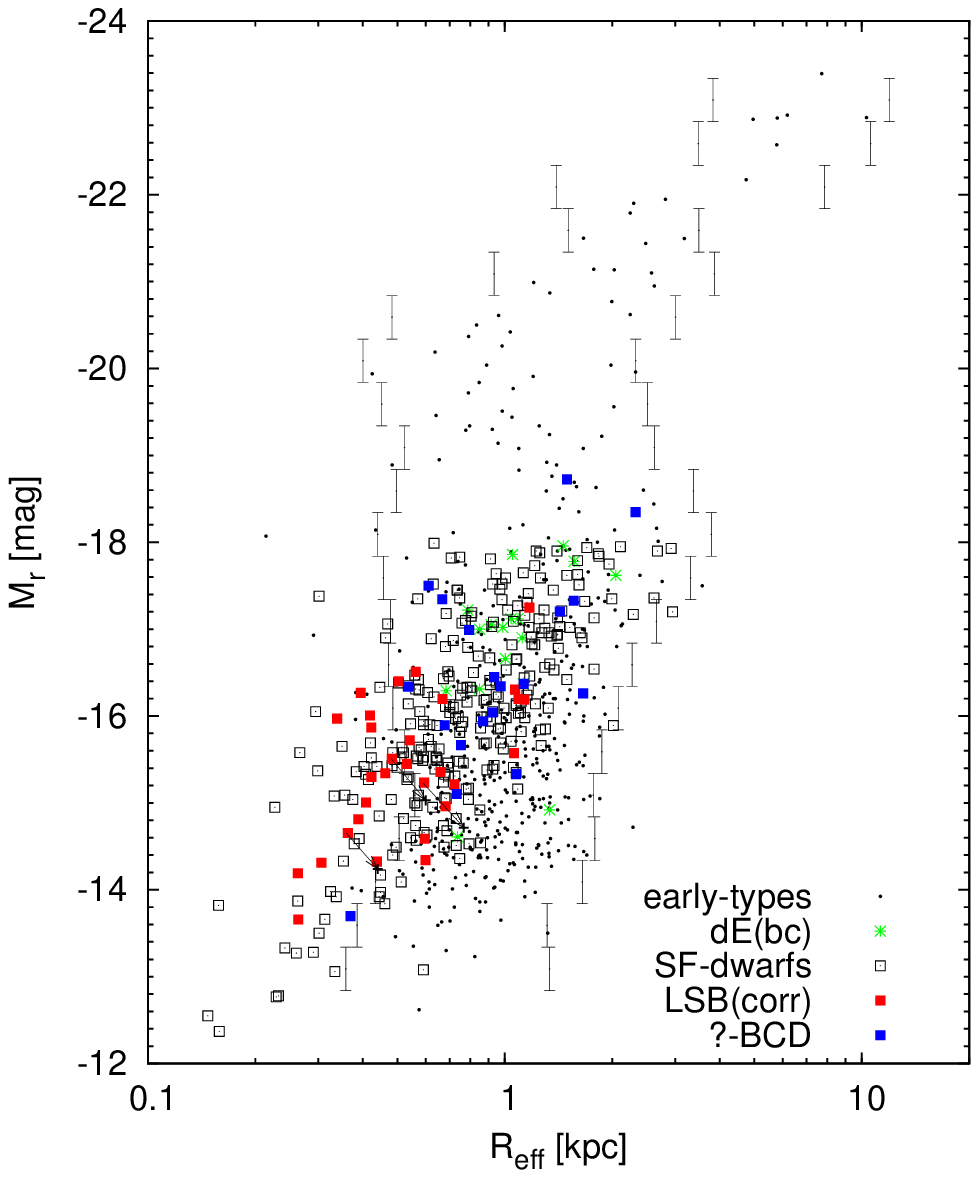} 
    \includegraphics[width=0.47\hsize]{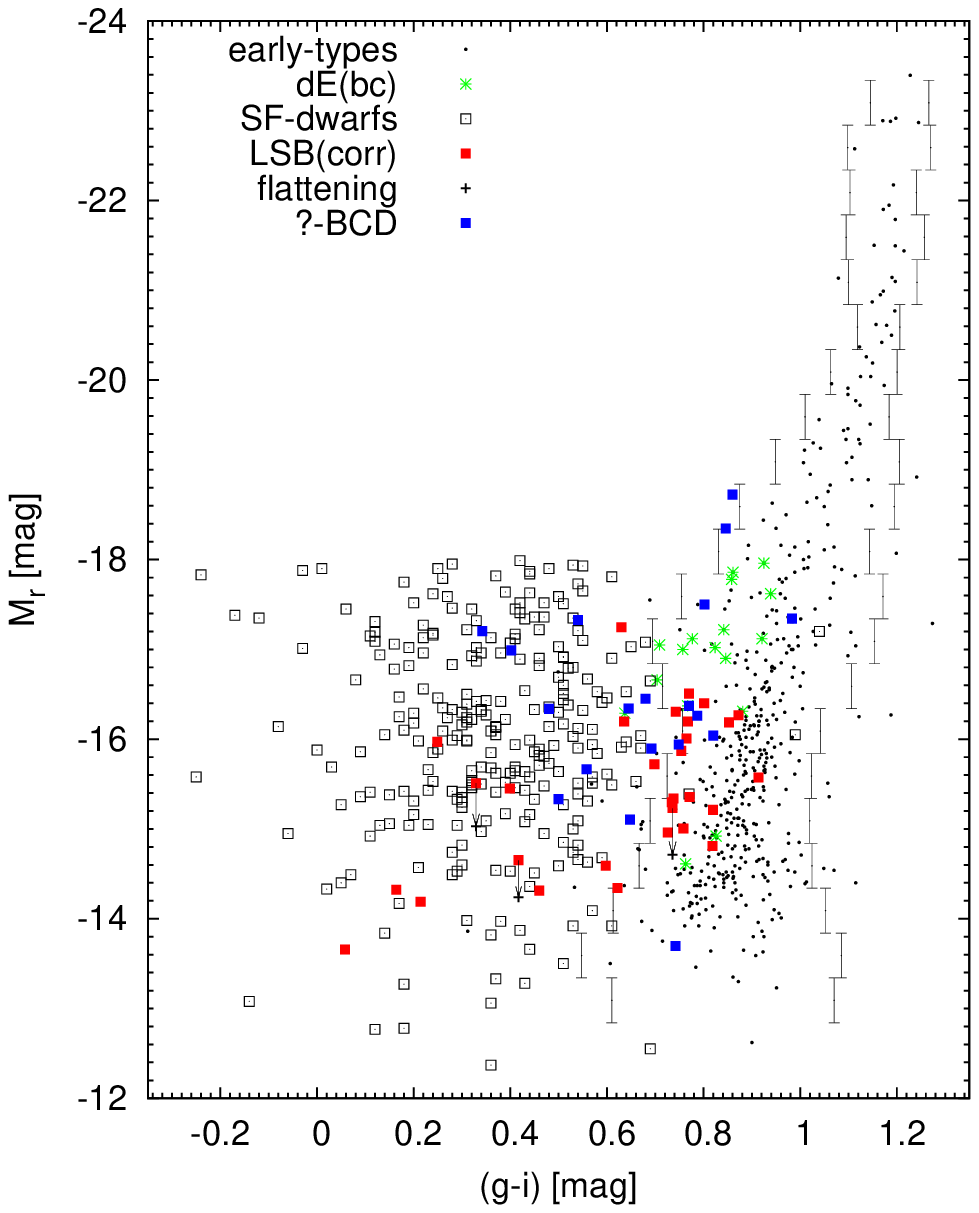}\\ 
    \includegraphics[width=0.49\hsize]{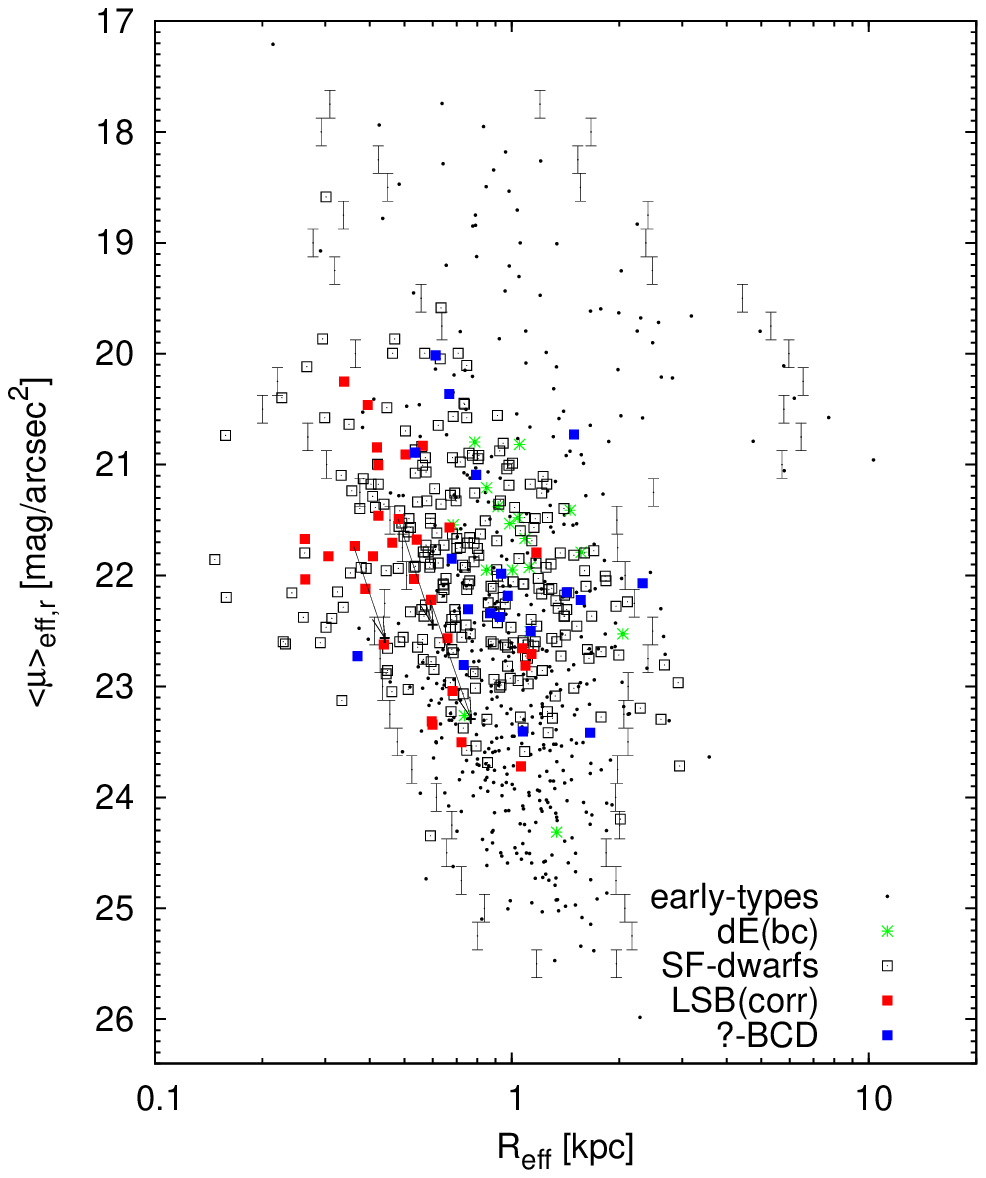}\\
      \caption{Parameter space of the LSB-components of our Virgo BCDs and a comparison sample of star-forming dwarf galaxies (SF-dwarfs) in {various} environments. The black squares correspond to the comparison sample of star-forming dwarf galaxies, while the other galaxy types are colour coded as in the previous figures.
              }
         \label{CMD-diplom}
   \end{figure*}

Figure~\ref{goldmine-velocity} shows the distance of the Virgo BCDs against the velocity ${v_{\rm LG}}$ with respect to the Local Group (cf.\ Fig.~5 of \citealt{Gavazzi99}), calculated via ${v_{\rm LG} = v_{\rm helio} + 220 \; km/s}$. The heliocentric velocity $v_{\rm helio}$ was given by GOLDMine. According to \citet{Gavazzi99} galaxies with ${v_{\rm LG} > 1900 \; km/s}$ belong to the M or W cloud with a distance of 32 Mpc.
As one can see, {three BCDs (VCC~0022, 0024, and 0274)} have a GOLDMine distance of 32 Mpc, but {have velocities either below or roughly at ${v_{\rm LG} = 1900 \; km/s}$}. Therefore, we do not assume a distance of 32 Mpc {for these galaxies}, but instead a distance of 16.5 Mpc.

Figure~\ref{goldmine-distance}  shows the (${M_r - R_{\rm eff}} -\left<\mu_{\rm{eff}}\right>$)-plots of the analysed galaxies with the GOLDMine distances. Interestingly, there are only three LSB-components of the BCDs that do not fall into the ETGs' $2\sigma$-plane, and these BCDs are classified as ``iI'' and ``iI,C''. One of these three BCDs {shows} a profile flattening of the LSB-component, shifting it {more to the $2\sigma$-region of the early types}. In the ${M_r - R_{\rm eff}}$-diagram, there are two {additional} BCDs outside the $2\sigma$-region of early-type dwarfs, {for which the above velocity criterion was not fulfilled and thus no greater distance was assigned.}

The comparison of the late-type dwarfs (BCD(-LSB) and irregular) and the early-type dwarfs reveals another interesting result. The LSB-components of the BCDs and the irregular galaxies form a continuous sequence, which {together shares} the same location as the early-type dwarf galaxies. This {implies that, structurally, the Virgo BCDs could evolve into early-type dwarfs with small radii once star formation ceases in them}. Additionally, Virgo irregulars could form the progenitor population of {future} early-type dwarfs with {large radii.}
Moreover, it should be pointed out that these findings are {dependent} on the distance of the galaxies. As seen in Fig.~\ref{Mreff}, there are more BCDs outside the $2\sigma$-region when a constant distance of 16.5 Mpc is adopted. {We point out that we have applied the individual GOLDMine distances only to the LSB-components of the primary BCD sample, in order to illustrate the effect on their values. In contrast, this would only have a weak effect on the population of early types: firstly, the vast majority of them are not located in the projected substructure regions that contain galaxies at significantly greater distances, and secondly, those that are located there would be shifted largely along the ${M_r - R_{\rm eff}}-$region that they already populate (cf.\ Table~\ref{tab:distances}). }

\subsection{Comparison with other BCDs}
\label{otherBCDs}

\begin{table*}
\caption{{Extremely metal-poor BCDs} from the study of \citet{Papaderos08}.}\tabularnewline
\label{tab-other-bcds}
\centering
\begin{tabular}{l c c c c c c c c c}
\hline\hline
Galaxy	&	${m_{\rm tot,g}}$	&	${M_{\rm tot,g}}$	&	${\mu_{\rm 0,g}}$	&	${\mu_{\rm eff,g}}$	&	${\left<\mu\right>_{\rm eff,g}}$	&	${R_{\rm eff,g}}$ 	&	D		&	${R_{\rm eff,g}}$	&	${R_{\rm eff,r}}$ \\
			&					&					&					&					&							&		[arcsec]		&	[Mpc]	& 	[kpc]				&	[kpc]\\ \hline
	J0133+1342	&	17.9				&	-15.0				&	22.1				&	34.7				&	23.22					&	4.63				&	37.8		&	0.85				&	0.89\\
	J1044+0353	&	17.2				&	-16.3				&	19.9				&	35.3				&	21.02					&	2.32				&	51.2		&	0.58				&	0.61\\
	J1201+0211	&	17.4				&	-13.3				&	22.2				&	32.5				&	23.32					&	6.10				&	14.0		&	0.41				&	0.44\\
	J1414$-$0208	&	18.1				&	-13.7				&	21.6				&	33.6				&	22.72					&	3.35				&	23.0		&	0.37				&	0.39\\
	J2230$-$0006	&	17.0				&	-15.0				&	19.8				&	33.8				&	20.92					&	2.43				&	24.9		&	0.29				&	0.31\\
	J2302+0049	&	18.6				&	-17.0				&	20.8				&	37.4				&	21.92					&	1.84				&	134.6	&	1.20				&	1.26
\\\hline                                   
\end{tabular}
\end{table*}

Table~\ref{tab-other-bcds} shows the values for a sample of BCDs from \citet{Papaderos08} (hereafter P08) from the SDSS. All given values of P08 were measured in the SDSS g-band and for a first order approximation we assume that ${R_{\rm eff}}$ is equal in the g and r filters.
As one can see, the effective radii of the P08 sample are also very compact with an average of ${\left<R_{\rm eff,g}\right> = 0.62 \;  kpc }$, but with a large scatter. 

Since the BCDs of our study are located in the Virgo cluster, one may ask whether they {deviate from BCDs in field and group environments}. To investigate this question a large sample of $340$
emission-line galaxies in {various} environments (unpublished study by H. T. Meyer, R. Kotulla, P. Papaderos, Y. Izotov, N.G. Guseva and K.J. Fricke) was used. The galaxies were spectroscopically selected, therefore, also non-dwarf galaxies are included in the initial sample. To include only star-forming \emph{dwarf} galaxies, a g-band magnitude cutoff of $M_{\rm cutoff} = -18.0$\,mag was applied,{ resulting in the sample that we use here for comparison}. 
This sample includes BCDs as well as other types of star-forming (dwarf)\footnote{The applied distances were simply converted from SDSS spectroscopic redshifts, which therefore do not account for possible peculiar velocities, and thus cause uncertainties in the absolute magnitudes and radii.} galaxies. 

To all galaxies of this {comparison sample} the same photometric analysis (derivation of profiles and decomposition) was applied. The CMD of the LSB-components of the Virgo BCDs and {the comparison sample} is shown in Fig.~\ref{CMD-diplom}. The LSB-components of the Virgo BCDs tend to be redder than those of the comparison galaxies. This could {indicate an} influence of the denser cluster environment on the Virgo BCDs, making them redder due to the removal of gas. In the {structural} diagrams of Fig.~\ref{CMD-diplom}, {one can see that the locus of the Virgo BCDs nearly completely overlaps with that of the comparison sample of star-forming dwarf galaxies in the field. The fact that the latter extends to larger effective radii and brighter magnitudes than the Virgo sample may reflect a selection effect.
An increase in the sample size of \emph{cluster} BCDs would therefore be desirable for future investigations.}

\subsection{Evolutionary connection between different galaxy types}
\label{evolution}

Various observational and theoretical lines of evidence suggest that the evolutionary pathways 
of late-type dwarfs in a cluster environment have to significantly differ from those in the field.
 {Several} mechanisms for starburst ignition {in the} initial stages of the interaction of a dwarf 
with the ICM can be envisaged, for example triggered gas collapse by the external ICM pressure (as opposed
to gas collapse driven by self-gravity), or strong dissipation and gas cooling in large-scale shocks, 
followed by collective star formation.
For such reasons, it is not even sure that the compact structure of the LSB-component -- 
a typical characteristic of field BCDs -- has to be common in
systems classified as BCDs in galaxy clusters. For example, external agents (see above) could ignite a 
BCD-typical starburst in a genuine (that is, relatively diffuse) irregular. Alternatively, an externally 
triggered destabilisation and inflow of gas could eventually lead to an adiabatic contraction 
of the LSB-component \citep{Papaderos96b}, and the opposite may happen after complete removal of the dwarf gas halo,
as it sinks deeper into the hostile ICM environment.

Even in the cluster periphery, late-type dwarfs have undergone interaction with  
the hot ICM and other nearby galaxies for a few hundreds of Myr, a time span on the order of their dynamical 
timescale. {In summary, in a late-type dwarf plunging into the ICM, the starburst that makes it appear as BCD} may have a 
different origin than starbursts in the general population of field BCDs. 
Consequently, one should be cautious when comparing structural and integral photometric properties 
 of cluster-BCDs with literature data for ordinary
BCDs.

As our understanding of the early interaction of late-type dwarfs with the cluster environment is still poor, 
it would be speculative to generalise conclusions drawn from our Virgo BCD sample to the BCD population
as a whole.
Nevertheless, the following discussion concentrates on a {nearly complete} and 
well-selected sample of cluster BCDs, thereby minimising  {the potential dilution of any trends that could be caused when combining objects from different environments}.

The evolutionary scenario of \citet{Davies88} describes how a dwarf irregular galaxy can ignite a starburst and become 
a BCD due to infalling gas from a surrounding reservoir.
After several starburst phases and non-starbursting irregular-phases, the galaxy evolves to a passive dwarf galaxy.
{If this evolution was possible in the Virgo cluster outskirts, we should observe a similar structure of BCDs (the starburst phase) and early-type dwarfs (the end product).}
While this may serve as a possible evolutionary path from the faint end of the late-type sequence of 
\citet{KormendyBender2012} to the faint end of the early-type sequence, we note that early-type and late-type galaxies in today's clusters have spent, \emph{on average}, a large fraction of their lifetime in different environments \citep{Lisker13}.

Observational studies of BCDs by \citet{Drinkwater91}\footnote{It should be pointed out that the findings of \citet{Drinkwater91} {were based on a particular} sample of the most compact BCDs in Virgo.} and \citet{Papaderos96b} {reported} that BCDs are very compact objects as compared to the early-type dwarfs. They concluded that only the most compact {of the latter} may be related to the BCDs.
An even more extreme conclusion is reached by \citet{Drinkwater96}, who state that ``no blue star-forming progenitors of dE galaxies'' exist in the Virgo cluster down to an absolute $B$-magnitude of M$_B \approx-14$ mag.
At first view, these findings {seemed} to be supported by our photometric measurements for the galaxies' entire light: Fig.~\ref{all-values-BCDs} (top left panel) shows that the majority of BCDs are more compact than even the $2\sigma$-region of early-type dwarfs at a given magnitude.

However, if the gas reservoir that is immediately available to a BCD's starburst region is limited, thus restricting the starburst phase to {a duration of some 100 Myr or less} \citep{Thuan91, Thornley00}, a population of non-starbursting counterparts must exist \emph{within the same spatial volume}. Assuming a velocity of 1000 km/s, a BCD could only move some 100 kpc during the starburst phase, i.e.\ some tenths of the Virgo cluster's virial radius --- which { would mean} that the outskirts of the Virgo cluster should hold {a substantial number of} non-starbursting counterparts to BCDs.\footnote{{This does not necessarily apply to BCDs that are located at greater distances than the main cluster, which, however, is the minority (Fig.~\ref{goldmine-velocity}).}}
How do these look like?

\begin{figure}[ht!]
\centering
\includegraphics[width=\hsize]{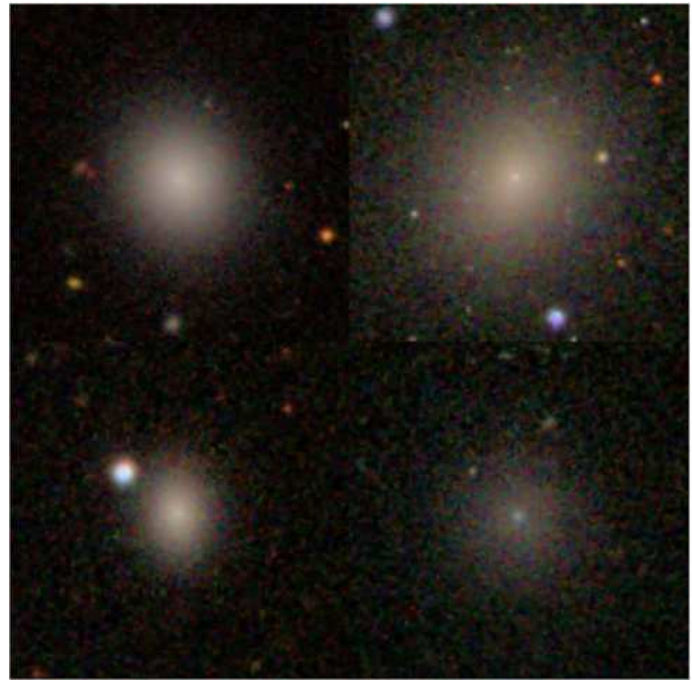}
\caption{{Possible descendants of BCDs (left) and of irregulars
    (right) in the Virgo cluster. The left panels show a bright (top)
    and a faint (bottom) early-type dwarf that fall into the
    structural parameter range of the LSB-components of Virgo BCDs,
    i.e.\ they are more compact than the average early-type
    dwarf. The right panels show a bright (top) and faint (bottom)
    early-type dwarf that fall into the range of the Virgo irregulars,
    i.e.\ they are more diffuse than the average early-type dwarf.
All images have a width of $80"$, corresponding to $6.3$\,kpc at a
distance of $16.5$\,Mpc.
The galaxies and their parameters are,
clockwise from top left:
VCC~0033 ($M_r = -16.85$\,mag, $R_{\rm eff} = 0.73$\,kpc),
VCC~0510 ($M_r = -17.00$\,mag, $R_{\rm eff} = 1.64$\,kpc),
VCC~1565 ($M_r = -15.49$\,mag, $R_{\rm eff} = 1.40$\,kpc), and
VCC~0068 ($M_r = -15.25$\,mag, $R_{\rm eff} = 0.53$\,kpc).
}
}
\label{descendant}
\end{figure}

To investigate this, we determined the structural properties of the underlying LSB-component of the Virgo BCDs, i.e.\ without the contribution of the starburst region. The results already show much more overlap with the early-type dwarfs in the parameter space of structure and also colour (see Figs.~\ref{CMD} and \ref{Mreff}). When we also take the different distances of the various parts of the Virgo region into account, the LSB-components turn out to fit remarkably well to the early-type dwarf population with small effective radii (Fig.~\ref{goldmine-distance}). 
This is in agreement with the conclusions of \citet{Micheva13} from deep imaging of a sample of luminous BCDs and intermediate-mass galaxies.
The left {panels} of Fig.~\ref{descendant} show {examples for} such early-type dwarfs that fall within the locus of the LSB-components:
these may look like the future ``red and dead'' descendants of BCDs. {In contrast, the right panels show early-type dwarfs that are more diffuse than the LSB-components, and instead overlap with the irregulars.}

  \begin{figure}[ht!]
   \centering
   \includegraphics[width=\hsize]{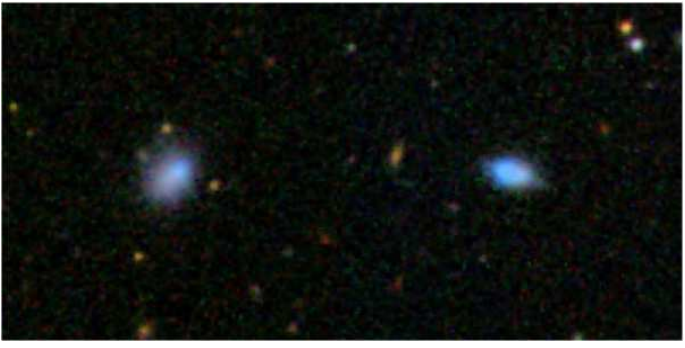}   
      \caption{{The two BCDs (left: VCC~0410, right: VCC~1313) that remain more compact than the early-type dwarfs even after applying the individual distances of GOLDMine. These objects lie outside of the $2\sigma$-region of early-type dwarfs in the $\left<\mu_{\rm{eff}}\right>-{R}_{\rm{eff}}$ diagram of Fig.~\ref{goldmine-distance}.
Both images have a width of $80"$, corresponding to $6.3$\,kpc at a
distance of $16.5$\,Mpc.
}
}
         \label{bcds-outsiders}
   \end{figure}

As for the {two} objects (Fig.~\ref{bcds-outsiders}) that remain more compact than the {$2\sigma$-region of early types in the $\left<\mu_{\rm{eff}}\right>-{R}_{\rm{eff}}$ diagram (Fig.~\ref{goldmine-distance}), their LSB-component colours are also bluer than the $2\sigma$-region of early types. VCC~1313 has hardly any discernible LSB-component within our detection limits. This galaxy seems }
to follow
the literal meaning of the term ``blue compact dwarf'' (see Fig.~\ref{bcds-outsiders}, right panels), while most of
the other {Virgo} BCDs rather have the appearance of a {(sometimes faint)} early-type dwarf or irregular with
additional starburst. The non-detection of a red LSB-component down to
low surface brightness levels for some ``extreme'' BCDs was already
reported by \citet{Drinkwater91}: two BCDs in their study had no
surrounding low-surface-brightness envelopes detected to a limit of
${\mu(V) = 27 \; {\rm mag/arcsec^2}}$. Therefore, they would be classified as i0-BCDs. For all of these, the Next Generation Virgo Cluster Survey \citep[NGVS,][]{Ferrarese12} will provide unprecedented insight into their outer structure and older stellar population, if existent.

\begin{figure}[ht!]
   \centering
\includegraphics[width=\hsize]{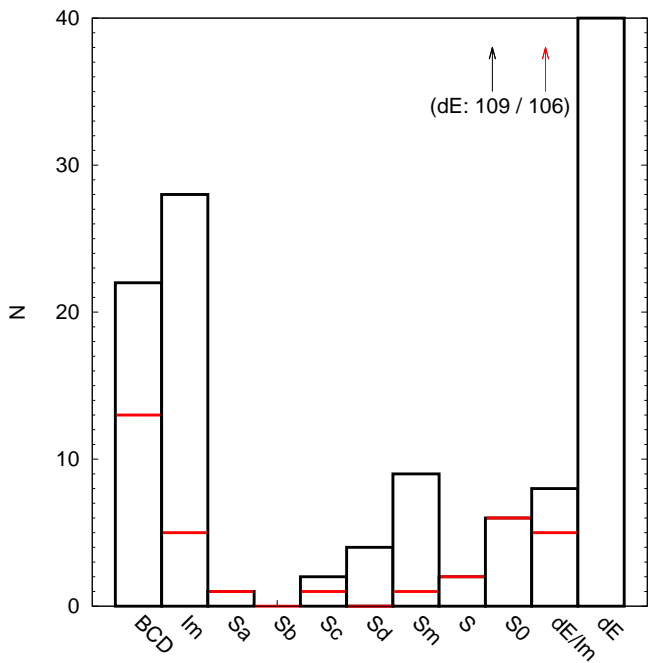}
\caption{{Histogram of morphological types} in the outer region of Virgo (${D_{\rm M87} > 0.6 \; }$Mpc). A cutoff of ${M_r > -16.5 \; mag}$ and ${\left<\mu_r\right>_{\rm eff,r} < 23.0 }\;$mag/arcsec$^2$ was applied to restrict the parameter region to the locus of most of the LSB-components {(see Fig.~\ref{Mr-mu}), using a distance of $16.5$\,Mpc for all galaxies. For the red histogram lines, an additional colour cutoff of ${(g-i)> 0.7}$ mag was applied}. The histogram of the early-type dwarfs (labelled ``dE'') also includes the dE(bc)s. {Candidate BCDs are included with their respective first-mentioned class (e.g.\ ``Sm / BCD''). The numbers are obtained from the full sample of Virgo cluster galaxies that was used in \citet{Weinmann11}, based on the VCC and adopting a faint magnitude limit of $18$\,mag in the $B$-band. For the galaxies that could not be analysed with the SDSS, we use estimations for $M_r$ as described in the appendix of \citet{Weinmann11}, and adopt the average surface brightness and colour of the respective class.}
}
\label{outer-virgo}
\end{figure}

Are early-type dwarfs the only morphological type that qualifies as non-starbursting counterparts of BCDs? Figure~\ref{outer-virgo} shows the number distribution of galaxies in the outer region of the Virgo cluster that fall in the same parameter space of magnitude and mean surface brightness as the {majority of the} LSB-components of the BCDs,
displayed separately for blue and red ($g-i>0.7$) galaxies.
{Among} blue galaxies, late-type spirals and especially irregular galaxies {\citep{Sanchez08}} form a population of significant size. However, the majority of the LSB-components of BCDs is already in the red colour regime {(red histogram in Fig.~\ref{outer-virgo})}, where the early-type dwarfs clearly outnumber all other galaxy types.

\citet{Vaduvescu06} concluded 
 from a near-infrared analysis of Virgo cluster galaxies, decomposed with a different scheme than the one used in this study,
that ``BCDs and dIs are similar structurally'', which is incompatible with our 
conclusions 
 (see Fig.~\ref{Mreff}). We measure irregulars to be systematically more extended than the LSB-components of BCDs at a given magnitude, hence only a small fraction of the irregulars would qualify as non-starbursting BCDs.
Vaduvescu et al.\ fitted structural parameter relations to irregulars, and claimed consistency of the BCDs with these relations. Their Figs.~4 and 5 show large scatter among their BCD sample, with 
nine of the 18 BCDs having substantially smaller radii than the relation of irregulars.
Among the five BCDs that lie significantly above the {relation of irregulars}, the three largest ones have been classified as ambiguous types between Sm, irregular and BCD in the VCC, and are more similar to {Sm/Im} by visual inspection of SDSS colour images. 
This may provide an explanation for the discrepant findings. 

{Indications that the Virgo BCDs have already experienced some environmental influence are given by the redder LSB-components of the Virgo BCDs as compared to a large BCD sample from various environments (Sect.~\ref{otherBCDs}), as well as by the fact that the integrated colours of many Virgo BCDs are not dominated by the starburst component (Fig.~\ref{all-values-BCDs}), consistent with their visual appearance but unlike typical field BCDs (see Sect.~\ref{sec:intro}). Due to these differences, we cannot be certain that the commonly assumed short duration of a BCD's starburst phase also holds for the Virgo BCDs. \emph{If} their starbursts are indeed
} relatively short, then many of the early-type dwarfs in the Virgo outskirts with similar magnitudes must be the non-starbursting counterparts to the BCDs. Unless we are witnessing all BCDs in their very last starburst phase, which seems unlikely, this 
would mean that \emph{a number of early-type dwarfs must be able to re-ignite a starburst}.

At first glance, this may seem consistent with the recent or even ongoing star formation in the dE(bc)s of the Virgo cluster \citep{Lisker06b}, as well as with the interpretation of \citet{Hallenbeck12} that gas detection in some Virgo early-type dwarfs could imply re-accretion of gas. However, {\citet{DeLooze13} recently reported from a {\it Herschel} far-infrared analysis that BCDs contain on average more dust than early-type dwarfs with residual star formation (``transition-type dwarfs''), and}
we find that the dE(bc)s are on average less compact and brighter than the LSB-components of the BCDs (see Fig.~\ref{Mreff}).
Future high-sensitivity {comparative studies of the stellar and gas distribution and dynamics} in those galaxy types are needed {for a conclusive interpretation of their possible evolutionary links}.

\begin{acknowledgements}
The authors would like to thank the anonymous referee for helpful suggestions.
HTM, TL, and JJ were supported within the framework of the Excellence
Initiative by the German Research Foundation (DFG) through the
Heidelberg Graduate School of Fundamental Physics (grant number GSC
129/1). HTM was supported by the DFG through grant LI 1801/2-1.
JJ acknowledges support by the Gottlieb Daimler and Karl Benz
Foundation.
PP is supported by Ciencia 2008 Contract, funded by FCT/MCTES 
(Portugal) and POPH/FSE (EC). He also acknowledges support by 
the Funda\c{c}\~{a}o para a Ci\^{e}ncia e a Tecnologia (FCT)
under project FCOMP-01-0124-FEDER-029170 
(Reference FCT PTDC/FIS-AST/3214/2012), funded by FCT-MEC (PIDDAC) 
and FEDER (COMPETE).

    Funding for the SDSS and SDSS-II has been provided by the Alfred P. Sloan Foundation, the Participating Institutions, the National Science Foundation, the U.S. Department of Energy, the National Aeronautics and Space Administration, the Japanese Monbukagakusho, the Max Planck Society, and the Higher Education Funding Council for England. The SDSS Web Site is http://www.sdss.org/.
    The SDSS is managed by the Astrophysical Research Consortium for the Participating Institutions. The Participating Institutions are the American Museum of Natural History, Astrophysical Institute Potsdam, University of Basel, University of Cambridge, Case Western Reserve University, University of Chicago, Drexel University, Fermilab, the Institute for Advanced Study, the Japan Participation Group, Johns Hopkins University, the Joint Institute for Nuclear Astrophysics, the Kavli Institute for Particle Astrophysics and Cosmology, the Korean Scientist Group, the Chinese Academy of Sciences (LAMOST), Los Alamos National Laboratory, the Max-Planck-Institute for Astronomy (MPIA), the Max-Planck-Institute for Astrophysics (MPA), New Mexico State University, Ohio State University, University of Pittsburgh, University of Portsmouth, Princeton University, the United States Naval Observatory, and the University of Washington.

\end{acknowledgements}


\begin{appendix}
\section{Error estimation}
\label{appendix}

To obtain a better insight into the errors of the derived parameters of the BCDs, we created a set of artificial BCDs. To mimick the real Virgo BCDs, we combined the light distribution from two components, representing the LSB and starburst component. Firstly, the LSB-components were divided into a bright version (${M_{\rm r,LSB} = -17.0 \;  mag}$), with half-light semi-major axes of ${a_{\rm hl} = 7.6 \;  arcsec}$ and ${a_{\rm hl} = 15.1 \;  arcsec}$, and a faint version (${M_{\rm r,LSB} = -15.0 \;  mag}$), with ${a_{\rm hl} = 6.3 \;  arcsec}$ and ${a_{\rm hl} = 12.6 \; arcsec}$. Secondly, two different S\'ersic indices $n$ were used, with $n=1.0$ (pure exponential) and $n=1.3$. We chose three different axis ratios ($b/a=0.5, 0.7, 0.9$) and two position angles (p.a.$=60, 90$) for the LSB component; the starburst component was chosen to be circular.

To mimick the additional outer tail of the LSB-component of some of the real Virgo-BCDs (see section~\ref{tail-section}) at larger radii, we optionally added an exponential outer tail to the exponential LSB-component, having the same axis ratio and position angle. The tail has twice the half-light semi-major axis of the LSB-component, and can be weak (${\Delta m = 1.2  \;}$  mag fainter in total brightness) or strong (${\Delta m = 0.75  }$\;  mag). Note that we did not add a tail to the LSB component with $n=1.3$, since our intention was to test whether our approach to assume an exponential shape for the LSB component's analysis, and to account for a possible outer tail, would be able to approximate the $n=1.3$ case reasonably well.

For the starburst component, two different effective radii ($2.0$ and $4.0$\; arcsec) were used to account for the various extents of the starbursts. We created one version in which the starburst is co-aligned with the LSB component, and another version in which it is offset by $0.5 {a_{\rm hl}}$, since a fraction of the real BCDs show a clear offset between starburst and LSB component (see Tab.~\ref{BCDsubtypes}).

With these different parameters we end up with 384 artificial BCDs. We first created noise-free images \emph{without} the starburst component, to determine the true total (two-Petrosian) magnitude ${ M_{\rm r,LSB}}$ of the LSB+tail component, as well as the true value of ${a_{\rm hl}}$, using the input axis ratios and position angles. We then created images with realistic noise and seeing characteristics, which were analysed in the same manner as the Virgo BCDs (see Section~\ref{photometry}). The magnitudes and radii derived with LAZY for the LSB+tail component were then compared to the true values, thereby yielding estimates for systematic errors of magnitude and \emph{relative} radius,
\begin{eqnarray}
 	\delta_{{M_{r}}}   & = & \left< {M_{r} - M_{\rm r,input}} \right>\\
 	\delta_{{R_{\rm eff}}} & = & \left< {R_{\rm eff} / R_{\rm eff,input}} \right>-1
\end{eqnarray}
as well as for their statistical errors (standard deviation) $\sigma_{{M_{r}}}$ and $\sigma_{{R_{\rm eff}}}$. Note that here and in the remainder of this section, we omit the subscript ``LSB'' from all quantities, for better readability.

\begin{table}
\caption{Mean values of the derived parameters for the LSB components of the artificial BCDs for different S\'ersic indices n. A possible outer tail was also taken into account for the LSB-component.}\tabularnewline
\label{tab:errors}
\centering
\begin{tabular}{l c c c c c}
\hline\hline
${M_{\rm r,input}}$ 	&	${n}$	&	$\left<{M_{\rm r} - M_{\rm r,input}}\right>$	&  $\sigma_{{M_r}}$	& $\left<{ R_{\rm eff} / R_{\rm eff,input}}\right>$	& $\sigma_{{R_{\rm eff}}}$	\\ \hline
	$-$17.0				&	1.0		&		$-$0.03					& 	0.08			&	1.01							&	0.06			\\	
	$-$17.0				&	1.3		&		$-$0.03					&	0.13			&	1.01							&	0.07			\\	
	$-$15.0				&	1.0		&		$-$0.15					&	0.10			&	0.89							&	0.05			\\	
	$-$15.0				&	1.3		&		$-$0.16					&	0.11			&	0.88							&	0.07
\\\hline                                   
\end{tabular}
\end{table}

\begin{table}
\caption{Final statistical ($\sigma_{{X}}$) and systematic ($\delta_{{X}}$) errors for the parameters of the LSB components, based on the analysis of artificial BCD images (Table~\ref{tab:errors}).}\tabularnewline
\label{tab:errors-mu}
\centering
\begin{tabular}{l c c c c c c}
\hline\hline
${M_{\rm r,input}}$ 	&	$\sigma_{{M_r}}$	 & $\sigma_{{R_{\rm eff}}}$	& $\sigma_{\left<\mu\right>_{\rm{eff}}}$	& $\delta_{{ M_{r}}}$ &  $\delta_{{R_{\rm eff}}}$	& $\delta_{\left<\mu\right>_{\rm{eff}}}$	 	\\\hline 
	$-$17.0				&	0.09				 &		0.06								&	0.16						&	$-$0.03			&	+0.01					& 	$-$0.01						\\	
	$-$15.0				&	0.10				 &		0.06								&	0.16						&	$-$0.15			&	$-$0.11					&	$-$0.41						
\\\hline                                   
\end{tabular}
\end{table}

Table~\ref{tab:errors} summarises the results, subdivided by LSB component brightness and profile shape. Based on this approach, the 
final systematic and statistical errors that we adopt for our BCD analysis are provided in Tab.~\ref{tab:errors-mu}, subdivided into bright and faint LSB components. We refrain here from differentiating between the different
 S\'ersic indices n and between the cases with/without a tail component, as the resulting values were very similar.

The mean effective surface brightness ${\left<\mu\right>_{\rm eff}}$ can be derived directly from ${M_{r}}$ and ${R_{\rm eff}}$:
\begin{equation}
 	{\left<\mu\right>_{\rm eff} = {m_{r}} + 2.5 \log \left( 2 \pi  \, {R_{\rm eff}}^2 \right)}
\end{equation}
Its statistical error was calculated assuming independent errors on magnitude and radius:
\begin{align}
 	\label{A:error-mu}	
	{\sigma_{\left<\mu\right>_{\rm{eff}}}}	& = \sqrt{	\left( \frac{\partial \left<\mu\right>_{\rm{eff}} } {\partial {R_{\rm eff}} }	\cdot 	\left(\sigma_{{R_{\rm eff}}}\cdot{R_{\rm eff}}\right) \right)^2 +	\left(		\frac{\partial \left<\mu\right>_{\rm{eff}}}{\partial {m_{r}}} 		\cdot 	\sigma_{{m_{r}}} \right)^2 }\\
		& =  \sqrt{ \left( 2.171 \cdot \sigma_{{R_{\rm eff}}} \right)^2		+  \left( \sigma_{{M_{r}}} \right)^2	}.
\end{align}
The systematic surface brightness error was calculated by
\begin{align}
 	\label{A:stat-error-mu}	
	{\delta_{\left<\mu\right>_{\rm{eff}}}}	& = \delta_{{M_{r}}} + 5 \cdot \log \left(1+ \delta_{{R_{\rm eff}}} \right).
\end{align}

The average differences between the input magnitudes and radii and the ones measured by LAZY are almost negligible for the bright galaxies ($0.03$\,mag / $1\%$) and moderate for the faint galaxies ($0.15$\,mag / $11\%$). However, both add up to a significant systematic effect on the surface brightness of $0.41$\,mag/arcsec$^2$ for the faint galaxies, while being negligible ($0.01$\,mag/arcsec$^2$) for the bright galaxies.
The statistical errors are reasonably small for all quantities, with $0.10$\,mag for ${M_r}$, $6\%$ for ${R_{\rm eff}}$, and $0.16$\,mag/arcsec$^2$ for ${\left<\mu\right>_{\rm eff}}$.

We point out that, due to our parameter setup, the fraction of artificial BCDs with inner flattening (see Section~\ref{innerflattening}) was much larger than for the real Virgo BCDs, where only three BCDs were fitted with a inner flattening. We therefore calculated the above quantities only from those BCDs without inner flattening, to avoid an overestimation of the errors. 
\end{appendix}

\end{document}